\documentclass[twocolumn,showpacs,eqsecnum]{revtex4b5}

\usepackage{graphicx}
\usepackage{dcolumn}
\usepackage{amsmath}
\usepackage{enumerate}

\begin{document}

\preprint{astro-ph/0101524}

\title{Sterile neutrino hot, warm, and cold dark matter}
\author{Kevork Abazajian}
\email{kabazajian@ucsd.edu}
\author{George M. Fuller}
\email{gfuller@ucsd.edu}
\author{Mitesh Patel}
\email{mitesh@physics.ucsd.edu}
\affiliation{Department of Physics, University of California, San Diego,
La Jolla, California 92093-0319}
\date{January 30, 2001}

\begin{abstract}

\bigskip 

We calculate the incoherent resonant and non-resonant scattering
production of sterile neutrinos in the early universe. We find ranges
of sterile neutrino masses, vacuum mixing angles, and initial lepton
numbers which allow these species to constitute viable hot, warm, and
cold dark matter (HDM, WDM, CDM) candidates which meet observational
constraints.  The constraints considered here include energy loss in
core collapse supernovae, energy density limits at big bang
nucleosynthesis, and those stemming from sterile neutrino decay:
limits from observed cosmic microwave background anisotropies, diffuse
extragalactic background radiation, and $^6$Li/D overproduction.  Our
calculations explicitly include matter effects, both effective mixing
angle suppression and enhancement (MSW resonance), as well as quantum
damping.  We for the first time properly include all finite
temperature effects, dilution resulting from the
annihilation or disappearance of relativistic degrees of freedom, and the
scattering-rate-enhancing effects of particle-antiparticle pairs
(muons, tauons, quarks) at high temperature in the early universe.
\end{abstract}


\maketitle

\section{Introduction}

In this paper we calculate the non-equilibrium resonant and
non-resonant production of sterile neutrinos in the early universe and
describe cosmological and astrophysical constraints on this dark matter
candidate. Depending on their masses and energy distributions, the
sterile neutrinos so produced can be either cold, warm, or hot dark
matter (CDM, WDM, HDM, respectively) and may help solve some
contemporary problems in cosmic structure formation. We can define
sterile neutrinos generically as spin-$1/2$, $SU(2)$-singlet particles
which interact with the standard $SU(2)$-doublet (``active'')
neutrinos $\nu_e$, $\nu_\mu$, and $\nu_\tau$, solely via ordinary mass
terms. Singlet neutrinos with masses $\sim10^{12} {\rm\,GeV}$ arise
naturally, for example, in ``see-saw'' models of neutrino mass in
grand unified theories (GUTs) \cite{seesaw}. Recent solar,
atmospheric, and accelerator neutrino oscillation experiments
\cite{superk,imb,davis,solar,lsnd}), however, imply the existence of
four light neutrino species with masses $\lesssim 10{\rm\,eV}$, only
three of which can be active, on account of limits on the invisible
decay width of the $Z^0$ boson \cite{pdg}. The remaining neutrino must
be sterile. The existence of multiple generations of quarks and
leptons in the standard model (SM) of particle physics, as well as many
independently-motivated extensions of the SM, imply that there are
additional, more massive sterile neutrinos with couplings to active
neutrinos currently beyond direct experimental reach. We describe
herein the cosmological implications of these heavier sterile
neutrinos.

Some early constraints on massive sterile neutrino production in the
early universe were derived in Refs.~\cite{fm,kt}. The first
analytical estimates of the relic sterile neutrino abundance from
scattering-induced conversion of active neutrinos were made by
Dodelson and Widrow \cite{dw}. They assumed a negligible primordial
lepton number, or asymmetry, so the neutrinos are produced
non-resonantly. They found that sterile neutrinos could be a 
WDM candidate with interesting consequences for galactic and large
scale structure. A process that can create sterile neutrino dark
matter with a unique energy distribution was proposed by Shi and
Fuller \cite{cooldm}. In that work a non-vanishing initial primordial
lepton number gives rise to mass-level crossings which enhance sterile
neutrino production, yielding dark matter with an energy spectrum
which is grossly non-thermal and skewed toward low energies. In
particular, the average energy is significantly less than that of an
active neutrino, and the neutrinos suppress structure formation on
small scales and behave like CDM on large scales ({\it i.e.}, they
behave as WDM or ``cool DM'').

At no previous time has there been a greater need for a more
comprehensive study of the cosmological and astrophysical consequences
of active-sterile neutrino mixing. The continuing influx of data from
neutrino experiments and new observations of galaxy cores and clusters
demand a more detailed understanding of the physics of sterile
neutrino dark matter production and more sophisticated calculations of
their relic abundances. 

We examine these experimental and observational issues in
Secs.~\ref{neutanomalysect} and \ref{structformsect}, respectively.
In Sec.~\ref{leptnumbsect} we discuss the present limits on primordial
lepton asymmetries and possible dynamical origins of these
asymmetries. We attempt to unify the perspectives of previous work, so
we take the initial lepton asymmetry of the universe to be a free
parameter within the rather generous bounds set by experiment.
After reviewing the physics of neutrino oscillations and matter-affected
neutrino transformation in Sec.~\ref{neutmixingsect}, we compute the
consequences of pre-existing lepton asymmetries for sterile neutrino
dark matter scenarios in Secs.~\ref{boltzsect} and
\ref{leptressect}. In Sec.~\ref{boltzsect}, we give the Boltzmann
equations governing the non-equilibrium conversion of active neutrinos
into sterile neutrinos and solve them in the limit of non-resonant
conversion, a limit which obtains when the lepton asymmetry is
sufficiently small, of order the baryon asymmetry. In
Sec.~\ref{leptressect}, we consider larger asymmetries and examine in
detail the physics of resonant neutrino transformation in the early
universe. For the non-resonant and several resonant cases we display
contours of constant relic density of sterile neutrinos in the plane
of neutrino mixing parameters. These results confirm and extend
previous work \cite{dw,cooldm}.
For both sets of calculations we also take into account
finite-temperature and finite-density effects for all three active
neutrino species; the dilution of sterile neutrino densities due to
heating of the photons and active neutrinos from the annihilation of
particle-antiparticle pairs before, during, and after the quark-hadron
(QCD) transition, and we allow for the enhanced active neutrino
scattering rates due to the increased number of scatterers in
equilibrium at high temperatures.
In Sec.~\ref{colldampsect} we calculate the collisionless damping
scales relevant for structure formation, scales which classify the
regions of dark matter parameter space as HDM, WDM, or CDM regions.
We consider in Sec.~\ref{cosmoconstraints} the limits on sterile
neutrino dark matter from the diffuse extragalactic background
radiation (DEBRA), cosmic microwave background (CMB), big bang
nucleosynthesis (BBN), and $^6$Li and D photoproduction.  Finally, in
Sec.~\ref{supernovaconstraints} we examine the implications of
active-sterile neutrino mixing in core-collapse (Type Ib/c, II)
supernovae. Conclusions are given in Sec.\
\ref{conclusions}. Throughout the paper we use natural units with
$\hbar=c=k_B=1$.

\section{Neutrino Anomalies}
\label{neutanomalysect}

Recent experiments have provided data indicating evidence for new
neutrino physics.  The most significant recent evidence is the
Super-Kamiokande Collaboration's statistically convincing result
\cite{superk}, verifying previous measurements \cite{imb}, of a
suppression of the atmospheric $\nu_\mu/\bar\nu_\mu$ flux. The most
persuasive Super-Kamiokande evidence for neutrino oscillations is the
measured zenith angle dependence of the $\nu_\mu/\bar\nu_\mu$ flux, an
observation fit most simply by maximal $\nu_\mu \rightleftharpoons
\nu_\tau$ mixing in vacuum, with vacuum mass-squared difference
$\delta m^2 \sim 3\times 10^{-3} {\rm\,eV}^2$. 

On another front, the ground-breaking observations by the Homestake
Collaboration found a solar neutrino flux far below that predicted on
the basis of sophisticated solar models \cite{davis}. The solar
neutrino problem has an interesting possible solution through
matter-enhanced resonant conversion via the
Mikheyev-Smirnov-Wolfenstein (MSW) mechanism \cite{msw} or through
vacuum or ``quasi-vacuum'' oscillations \cite{beacom}.  Depending on
whether the solar solution involves two-, three-, or four-neutrino
mixing, the parameter space of neutrino mass-squared difference and
vacuum mixing angle are constrained differently \cite{solar}.

A third indication for neutrino oscillations comes from the Los Alamos
Liquid Scintillator Neutrino Detector (LSND) Collaboration's
observations of excess $\nu_e$ and $\bar\nu_e$ events in beams of
$\nu_\mu$ and $\bar\nu_\mu$, respectively. These have been interpreted
as evidence for neutrino oscillations in the $\nu_\mu\rightarrow\nu_e$
and $\bar\nu_\mu\rightarrow\bar\nu_e$ channels \cite{lsnd}. The
Karlsruhe Rutherford Medium Energy Neutrino (KARMEN) experiment probes
the same channels but does not see evidence for neutrino oscillations
\cite{karmen}. A joint analysis of the LSND and KARMEN data has found
that there are regions of neutrino mixing parameter space consistent
with both experiments' results \cite{eitel}.  

Efforts have been made to embed the above neutrino oscillation
solutions within a three-neutrino framework
\cite{scott,cf,barger4,ackerpakvasa}. Leaving aside maximal vacuum
mixing of all three neutrino species \cite{scott}, these analyses
generally require the atmospheric neutrino oscillation length scale to
be the one associated with the short-base-line LSND experiment
\cite{cf,barger4}, or the solar and atmospheric solutions must be
built on the same mass difference scale \cite{ackerpakvasa}.

However, the zenith-angle dependence of the Super-Kamiokande
measurement requires the atmospheric neutrino oscillation length to be
much larger than the corresponding LSND scale \cite{superk,lsnd}. This
disfavors the first three-neutrino scheme. Additionally, the three
solar neutrino experimental modes presently available suggest an
energy dependence in the $\nu_e$ survival probability which is likely
inconsistent with the second three-neutrino scheme. Taking these two
length scales, and the results of global flux measurement fits for the
solar neutrino oscillation interpretation \cite{solar}, the three
different oscillation length and energy scales require three disparate
mass differences, which cannot be accommodated in a three-neutrino
framework. The CERN $e^+e^-$ collider LEP measurement of the $Z^0$
width indicates the number of active neutrinos with masses $<m_Z/2$ is
$3.00\pm 0.06$ \cite{pdg}, so the results are {\it prima facie}
evidence for a light sterile neutrino species.

A number of neutrino mass models can provide the masses and mixings
needed to accommodate all of the neutrino oscillation data
\cite{langacker,arkanigrossman,chunkim,romao,shadow,dienesarkani}.
For example, in some string theories, higher-dimensional operators,
suppressed by the powers of the ratio of some intermediate mass scale
and the string scale, can give the light and comparable Dirac and
Majorana masses necessary for appreciable active-sterile neutrino
mixing \cite{langacker}. In theories with light composite fermions,
several of the fermions may mix with standard model neutrinos, giving
light active and sterile neutrinos \cite{arkanigrossman}.  In the
minimal supersymmetric standard model (MSSM) with explicit $R$-parity
violation, a neutralino can provide the required mixing for the
atmospheric and solar neutrino problems \cite{chunkim,romao}. This
model can also supply a sterile or ``weaker-than-weakly'' interacting
particle that has a small mixing with one or more active neutrino
flavors.  A low-energy extension of the Standard Model with an $SO(3)$
gauge group acting as a ``shadow sector'' may result in a neutral
heavy lepton \cite{shadow}.  A model with several sterile neutrino
dark matter ``particles'' arises in brane-world scenarios. It invokes
bulk singlet fermions coupling with active neutrinos on our brane
\cite{dienesarkani}.  However, many models of bulk neutrinos as
sterile neutrino dark matter must be rather finely tuned and must
avoid several cosmological constraints \cite{afp1}.

Although these mass models have been proposed to account for the
existing neutrino anomalies, many of them already contain or can be
easily be extended to contain additional singlet states which also mix
with active neutrinos. As long as the new particles are sufficiently
massive and have sufficiently small mixings with active neutrinos,
they evade terrestrial constraints, and it is interesting and useful to
speculate on their cosmological and astrophysical consequences.  As we
show later in this paper, even apparently negligible active-sterile
neutrino mixing is sufficient to induce the production of sterile
neutrino dark matter in the early universe.

A specific model's viability in producing a sterile neutrino dark
matter candidate depends on whether the mass and mixing properties of
the candidate(s) with the active neutrinos lie within the range that
produces an appropriate amount of dark matter, and whether the
candidate is stable over the lifetime of the universe, does not
engender conflicts with observationally-inferred primordial light
element abundances, does not violate CMB bounds, and does not
contribute excessively to the DEBRA in photons
\cite{ressellturner}. Potential constraints on these scenarios also
may arise from deleterious effects associated with the neutrino
physics of core-collapse (Type Ib/c, II) supernovae (see, {\it e.g.},
Refs. \cite{super,super2,slaclectures} and
Sec.~\ref{supernovaconstraints}).  Constraints on massive sterile
neutrinos ($10{\rm\,MeV}\leq m_s\leq 100{\rm\,MeV}$) from the SN 1987A
signal and BBN were considered previously in Ref.\ \cite{dhrs2}.

The existence of massive sterile singlet neutrinos that mix with
$\nu_e$ has been probed in precision measurements of the energy
spectrum of positrons in the pion decay $\pi^+ \rightarrow e^+\
\nu_e$.  The best current limits are from Britton {\it et~al.},
\cite{britton}, which constrain the mixing matrix element $\vert
U_{ex}\vert^2 < 10^{-7}$ for sterile neutrino masses $50\,{\rm MeV} <
m_s < 130\,\rm MeV$.  These limits may be significantly improved in
future precision experiments \cite{behrpersonal}.  Less stringent
constraints from peak and kink searches exist for smaller $m_s$. In
addition, searches for decay of massive $\nu_s$ have yielded
constraints for $|U_{\mu x}|^2$, $|U_{\tau x}|^2$ \cite{nomad}, as
well as $|U_{ex}|^2$ \cite{pdg}.

\section{Structure Formation}
\label{structformsect}
 
Conflicts may have appeared between standard cold dark matter
theory and simulations and observations of large and small scale
cosmological structure.  Simulations predict about 500 small halos
({\it i.e.}, dwarf galaxies) with mass greater than $10^8\,M_\odot$
around a galaxy like the Milky Way, but only 11 candidates are
observed near the Milky Way \cite{dwarf1}, and only 30 in the local
group out to $\sim$1.5 Mpc \cite{dwarf2}. In other words, simulations
of the standard $\Lambda$-CDM model {\it may} predict more dwarf
galaxies than are seen (but see Ref.\ \cite{primack}). Also, even if
the dwarf halos are dark, their overabundance may hinder galactic disk
formation \cite{wolfex}.

Another potential problem with CDM simulations is the appearance of
singularities or ``cusps'' of high density in the cores of halos. The
observations of the innermost profiles of galaxy clusters are
ambiguous \cite{innerhalo}, but rotation curves of the central regions
of dark matter-dominated galaxies consistently imply low inner
densities \cite{lowdens}. Recent $N$-body calculations of the
nonlinear clustering of WDM models have found that enhanced
collisionless damping can lower halo concentrations, increase core
radii, and produce far fewer low mass satellites \cite{bode}.  Also,
the observed phase space density in dwarf spheroidal galaxies may
suggest a primordial velocity dispersion like that of WDM
\cite{hogandelcanton}.

A promising new constraint on the nature of dark matter may come from
study of the Ly$\alpha$ forest in the spectrum of high-redshift
quasars \cite{earlylya,croftsd}.  The structure of the Ly$\alpha$
forest at high redshifts has been used to constrain the contribution
of HDM and, therefore, the mass of the active neutrinos
\cite{crofthudave}. The constraints presented in Ref.\
\cite{crofthudave} and the Gerstein-Zeldovich--Cowsik-McLelland bound
\cite{cmbound} only pertain to fully populated active neutrino seas
with a thermal energy spectrum.  For example, if an active neutrino
has a mass $m_{\nu_\alpha}\sim 1\,\rm keV$ ($\alpha=\mu,\tau$)  and
the reheating temperature of inflation is low ($T_{\rm RH}\sim 1\,\rm
MeV$), then the corresponding active neutrino sea will not be fully
populated \cite{kawasaki,guidice1} and can be WDM \cite{guidicekolb}.
However, observations of the power spectrum of the Ly$\alpha$ forest
cannot only constrain HDM scenarios, but also may be able to test the
viability of WDM scenarios.  Since the absolute normalization of the
power spectrum is uncertain, the relative presence of power between
scales near $\sim\! 1\,\rm Mpc$ and near $\sim\! 100\,\rm kpc$ may
constrain the WDM scenarios that have a relatively large contrast in
power between these scales \cite{scottpersonal}.  In addition, the
large number of high-redshift quasars found by the Sloan Digital Sky
Survey (SDSS) \cite{sdss} can add considerably to the knowledge of the
Ly$\alpha$ power spectrum, particularly at the largest scales. The
combination of the galactic power spectrum from SDSS and the
Ly$\alpha$ power spectrum will produce even stronger constraints on
the behavior of dark matter on small scales.

\section{Primordial Lepton Asymmetry}
\label{leptnumbsect}

The lepton number or asymmetry of a neutrino flavor $\alpha$ is
defined to be
\begin{equation}
L_\alpha \equiv \frac{n_{\nu_\alpha} - n_{\bar\nu_\alpha}}{n_\gamma},
\end{equation} 
where $n_{\nu_\alpha}$ is the proper number density of neutrino
species $\nu_\alpha$, and $n_\gamma = 2\zeta(3)T^3/\pi^2 \approx
0.243T^3$ is the proper number density of photons at temperature
$T$. The lepton number(s) of the universe is (are) not well
constrained by observation. The best limits come from the energy
density present during BBN and the epoch of decoupling of the CMB
\cite{scherrer,KS,espokaj,hannlesg}. In fact, the best current bounds
on the lepton numbers come from the observational limits on the $^4$He
abundance, radiation density present at the CMB decoupling, and
structure formation considerations \cite{KS,espokaj,hannlesg}:
\begin{equation}
\label{limite}
-4.1\times{10}^{-2}\le L_{\nu_e}\le 0.79,
\end{equation}
\begin{equation}
\label{limitmu}
\vert L_{\nu_\mu,\nu_\tau}\vert \le 6.0 .
\end{equation}
The bound on positive $L_{\nu_e}$ is weaker than that for negative
$L_{\nu_e}$ since it is possible to combine the effects of
neutron-to-proton ratio ($n/p$) reduction of positive $L_{\nu_e}$ with
a large $L_{\nu_\mu}$ or $L_{\nu_\tau}$, which increases the expansion
rate and thus the $n/p$ ratio entering BBN.  This cancellation could
provide a neutrino ``degenerate'' BBN that could replicate not only
the primordial $^4$He, but also the $\rm D/H$ and $^7$Li abundances
predicted by standard BBN.  See Refs.\ \cite{KS,espokaj} for a further
discussion.  As stated, the limits (\ref{limite}), (\ref{limitmu})
depend on an assumption of a roughly Fermi-Dirac,
low-chemical-potential energy spectrum for each neutrino species. They
do not address neutrino mass and do not take account of potential
bounds stemming from closure ({\it i.e.}, age of the universe)
considerations. The above limits do not change significantly when
recent estimates of observationally-inferred primordial abundances are
employed \cite{espokaj}. In any case, it is clear that neutrino number
density asymmetries less than about $10\%$ of the photon number are
easily allowed. However, we will see that lepton asymmetries at this
level, or even several orders of magnitude smaller, could have a
significant and constrainable effect if there are massive sterile
neutrinos (see Sec. \ref{leptressect}).

An issue which arises whenever the lepton number(s) $L_{\nu_\alpha}$
differ from the baryon number $B$ (or $\eta = n_b/n_\gamma \approx
2.79\times 10^{-8} \Omega_b h^2 \sim 10^{-10}$, where $n_b$ is the
proper baryon number density and $\Omega_b$ is the baryon rest mass
closure fraction and $h$ is the Hubble parameter in units of $100\rm\,
km\, s^{-1}\, Mpc^{-1}$) concerns the process of baryogenesis.  For
example, electroweak baryogenesis predicts the equality of these
numbers $B-L=0$. Note that there also exist simple processes that
violate $B-L$ and create lepton and/or baryon number either through
the Affleck-Dine mechanism \cite{affleckdine} or through
non-equilibrium decays of a heavy Majorana particle
\cite{buchmuller}. In Ref. \cite{bplusl}, several natural scenarios
of producing a large lepton number and the observed small baryon
number were investigated. Furthermore, lepton number could arise
spontaneously through matter-enhanced active-sterile neutrino
transformation at energy scales below that of the baryogenesis epoch
\cite{fvamp,shi96}. Finally, it should be recognized that through
cancellation of lepton numbers we could have $B-L=0$ (where
$L=L_{\nu_e} + L_{\nu_\mu} + L_{\nu_\tau}$), while still having
significant lepton-driven weak potentials [see Eq.\ (\ref{L})] in the
early universe.

\section{Matter-Affected Neutrino Transformation}
\label{neutmixingsect}

\subsection{General framework}

Neutrino mixing phenomena arise from the non-coincidence of
energy-propagation eigenstate and the weak (interaction) eigenstate
bases. Eigenstates of neutrino interaction (flavor) include the
active neutrinos ($\nu_e$, $\nu_\mu$, $\nu_\tau$) which are created
and destroyed in the standard model weak interactions, as well as
sterile neutrinos ({\it e.g.}, $\nu_s$, $\nu_s^\prime$,
$\nu_s^{\prime\prime}$, ...) which do not participate in weak
interactions. Eigenstates of neutrino propagation are states of
definite mass and energy (or momentum) and evolve independently of
each other between weak interaction vertices. If the bases spanned by
these sets of eigenstates happen to coincide, then active and sterile
neutrinos propagate independently between interactions. In general,
however, the bases need not coincide, since the symmetries of the
standard model and its many proposed extensions do not require the
unitary transformation between the bases to be the identity
transformation. As a result, neutrinos can oscillate, or transform in
flavor, between interactions.

This physics applies for any number of neutrino flavors, including the
four (three active plus one sterile) which can accommodate the
neutrino experiments. In the environments we consider, active-active
mixing is suppressed due to the very similar matter effects for all
active species.  In general, the 4-neutrino evolution may have
multiple active species mixing with the sterile neutrino, or in
multiple-sterile scenarios, there may be mixing between the sterile
neutrinos.  In our analysis here, we consider the simplifying limit of
two neutrino (active-sterile) mixing to explore the basic physics of
sterile neutrino dark matter production.

In the case of two-neutrino mixing, the unitary transformation
between the bases can be written as
\begin{eqnarray}
|\nu_\alpha \rangle &=& \cos \theta |\nu_1\rangle + \sin \theta | \nu_2
 \rangle \nonumber \cr
|\nu_s \rangle &=& -\sin \theta |\nu_1\rangle + \cos \theta | \nu_2
 \rangle
\end{eqnarray}
where $|\nu_\alpha\rangle$ and $|\nu_s\rangle$ are active ($\alpha =
e,\mu,\tau$) and sterile neutrino flavor eigenstates, respectively,
and $|\nu_1\rangle$ and $|\nu_2\rangle$ are neutrino mass (energy)
eigenstates with mass eigenvalues $m_1$ and $m_2$, respectively. The
vacuum mixing angle $\theta$ parametrizes the magnitude of the mixing
(and, as we shall see, the effective coupling of the sterile neutrino
in vacuum). We choose all of the neutrino flavor and mass eigenstates
to be eigenstates of momentum with eigenvalue $p$. Then a mass-energy
eigenstate $|\nu_i\rangle$ $(i=1,2)$ develops in time and space with
the phase
\begin{equation}
e^{i\vec p_i\cdot \vec x} = e^{i({\bf p}\cdot{\bf x} - E_it)} = e^{i(px -
\sqrt{m_i^2 + p^2}t)} \approx e^{-ix m_i^2/2p},
\label{phaseeq}
\end{equation}
where $E_i=\sqrt{m_i^2+p^2}$ is the energy of the eigenstate; $p\equiv
|{\bf p}|$ is the magnitude of the proper momentum of the species. If
the neutrino mass eigenstates are relativistic so that $E_i\gg m_i$,
we have $E_i \approx p+m_i^2/2p$ and $x\approx t$, yielding the last
approximation in Eq.\ (\ref{phaseeq}).  (In this last approximation we
suppress the part of the evolution operator proportional to the trace,
as this simply gives an overall common phase to the states.)  In our
study of sterile neutrino production, the sterile neutrinos are always
relativistic during the epochs in which they were produced.

The difference of the squares of the vacuum neutrino mass eigenvalues
is, for example, $\delta m^2 = m_2^2-m_1^2$. We can follow the
evolution of a coherently propagating neutrino state
$|\Psi_\nu\rangle$ in either the mass-energy or flavor basis. In the
flavor basis, a Schr\"odinger-like equation describes how the flavor
amplitudes, $a_\alpha(x)=\langle \nu_\alpha|\Psi_\nu(x) \rangle$ with
$\alpha = e,\mu,\tau$ and $a_s(x)=\langle \nu_s|\Psi_\nu(x) \rangle$,
develop with time-space coordinate $x$:
\begin{widetext}
\begin{equation}
i \frac{d}{dx}
\begin{pmatrix} a_\alpha \cr a_s\end{pmatrix}
=
\left\{
\left(p+\frac{m_1^2+m_2^2}{4p}+\frac{V(x,p)}{2}\right) I
+\frac{1}{2}
\begin{pmatrix}
V(x,p)-\Delta(p)\cos2\theta & \Delta(p)\sin2\theta\cr
\Delta(p)\sin2\theta & \Delta(p)\cos2\theta-V(x,p)
\end{pmatrix}
\right\}
\begin{pmatrix} a_\alpha \cr a_s\end{pmatrix},
\end{equation}
\end{widetext}
where the first term is proportional to the identity and $\Delta(p)
\equiv \delta m^2/2p$. In the context of the early universe it is most
convenient to take $x=t$, the Friedman-Lemaitre-Robertson-Walker
coordinate time (age of the universe), while in supernovae we take $x$
to be position.  The weak potential $V(x,p)$ represents the effects of
neutrino neutral current and charged current forward scattering on
particles in the plasma carrying weak charge. In the early universe we
have $V(T,p)$, but in supernovae, $V$ also depends on position $x$.
We suppress the $x$ and $T$ dependence of $V$ in the rest of the
paper. For a review of these issues and neutrino astrophysics, see,
{\it e.g.}, Refs.\ \cite{haxtonbal,fullerchap,fuller87,nr}.

Neutrino mixing can be modified by the presence of a finite
temperature background and any asymmetry in lepton number.  The
oscillation length is
\begin{equation}
l_m = \left\{\Delta^2 (p) \sin^2 2\theta + \left[\Delta (p) \cos
2\theta - V^D - V^T(p)\right]^2\right\}^{-1/2}.
\end{equation}
The effective matter-mixing angle is
\begin{equation}
\sin^2 2\theta_m = \frac{\Delta^2 (p) \sin^2 2\theta}{\Delta^2 (p)
\sin^2 2\theta + \left[\Delta (p) \cos 2\theta - V^D -
V^T(p)\right]^2}.
\end{equation}
Matter effects have been separated into finite density and finite
temperature potentials, $V^D$ and $V^T (p)$.  

The finite density potential $V^D$ arises from {\it asymmetries} in
weakly interacting particles ({\it i.e}, nonzero lepton numbers), not
from non-zero total densities alone.  In general, the finite density
potential is \cite{smf,nr}
\begin{widetext}
\begin{equation}
V^D = 
\begin{cases}
\sqrt{2}G_F\left[2(n_{\nu_e}-n_{\bar{\nu}_e}) +
(n_{\nu_\mu}-n_{\bar{\nu}_\mu})
+ (n_{\nu_\tau}-n_{\bar{\nu}_\tau}) +
(n_{e^-}-n_{e^+}) - n_{n}/2\right] & \text{for\ } \nu_e\rightleftharpoons\nu_s,
\\
\sqrt{2}G_F\left[(n_{\nu_e}-n_{\bar{\nu}_e}) +
2(n_{\nu_\mu}-n_{\bar{\nu}_\mu}) +
(n_{\nu_\tau}-n_{\bar{\nu}_\tau}) - n_{n}/2\right] & \text{for\ }
\nu_\mu\rightleftharpoons\nu_s,
\\
\sqrt{2}G_F\left[(n_{\nu_e}-n_{\bar{\nu}_e}) +
(n_{\nu_\mu}-n_{\bar{\nu}_\mu}) +
2(n_{\nu_\tau}-n_{\bar{\nu}_\tau}) - n_{n}/2\right] & \text{for\ }
\nu_\tau\rightleftharpoons\nu_s.
\end{cases}
\label{generalvd}
\end{equation}
\end{widetext}

The thermal potential $V^T$ arises from finite temperature effects and 
neutrino forward scattering on the seas of thermally created particles
\cite{nr}:
\begin{eqnarray}
\label{vt}
V^T (p) = &-& \frac{8\sqrt{2} G_{\rm F} p_\nu}{3 m_{\rm Z}^2}
\left(\langle E_{\nu_\alpha} \rangle n_{\nu_\alpha} + \langle
E_{\bar\nu_\alpha} \rangle n_{\bar\nu_\alpha}\right) \cr &-&
\frac{8\sqrt{2} G_{\rm F} p_\nu}{3 m_{\rm W}^2} \left(\langle E_\alpha
\rangle n_\alpha + \langle E_{\bar\alpha} \rangle
n_{\bar\alpha}\right),
\end{eqnarray}
where $n_\alpha$ ($n_{\bar\alpha}$) is the proper number density of
leptons (anti-leptons) of flavor $\alpha$ and $\langle E_\alpha
\rangle\ (\langle E_{\bar\alpha}\rangle)$ is the average energy of the
lepton (antilepton), and $n_{\nu_\alpha}$ ($n_{\bar\nu_\alpha}$) and
$\langle E_{\nu_\alpha} \rangle\ (\langle E_{\bar\nu_\alpha}\rangle)$
are the proper number density and average energy of the neutrinos
(antineutrinos) of flavor $\alpha$. The second term in Eq.\ (\ref{vt})
must be included whenever the lepton of the same flavor as the active
neutrino in question is populated.

In this paper, we will assume that the {\it initial} neutrino
distribution functions are close to Fermi-Dirac black bodies, which
for occupation of differential interval $d p$ have the form
\begin{eqnarray}
\label{fnu}
dn_{\nu_\alpha} &\approx&
{\frac{n_{\nu_\alpha}}{T^3 F_2\left(\eta_{\nu_\alpha}\right)}}
{\frac{p^2 dp}{e^{E(p)/T-\eta_{\nu_\alpha}}+1}}\nonumber \\
&\approx& \left({\frac{n_\gamma}{4 T^3 \zeta\left(3\right)}}\right)
{\frac{p^2 dp}{e^{E(p)/T}+1}},
\end{eqnarray}
where $E(p) = (p^2 + m^2)^{1/2}$, and $E(p)\approx p$ in the
relativistic kinematics limit. In this expression $\eta_{\nu_\alpha}
=\mu_{\nu_\alpha}/T$ is the degeneracy parameter (chemical potential
divided by temperature) for neutrino species $\nu_\alpha$ and
$F_2(\eta)\equiv \int_0^\infty x^2dx/(e^{x-\eta}+1)$ is the
relativistic Fermi integral of order $2$ $[F_2(0)=(3/2)\zeta(3)]$.
The distribution function for a neutrino species $\alpha$ is 
\begin{equation}
f_\alpha(p,t) = 1/(e^{E(p)/T-\eta_{\nu_\alpha}}+1).
\label{distfunc}
\end{equation}
The last approximation in Eq.\ (\ref{fnu}) follows if we take the
neutrino degeneracy parameter to be zero. This is frequently a good
approximation for almost all of the range of lepton numbers which are
interesting for our purposes. However, it may not be valid over the
broader range of allowed lepton numbers given in Eqs.\ (\ref{limite})
and (\ref{limitmu}). For small lepton numbers $\eta_{\nu_\alpha}
\approx 1.46 L_{\nu_\alpha}$; in fact, whenever $\nu_\alpha$ has
relativistic kinematics, we can write
\begin{equation}
\label{Leta}
L_{\nu_\alpha} \approx
{\frac{1}{4\zeta(3)}}{\left\{{\frac{\pi^2}{3}}\eta_{\nu_\alpha}+{\frac{1}{3}}
\eta_{\nu_\alpha}^3\right\}}.
\end{equation}
This equation can easily be inverted to find
$\eta_{\nu_\alpha}(L_{\nu_\alpha})$.

In the early universe, for temperature ranges where the number of
degrees of freedom is constant, the time-temperature relation is a
simple power law and the quantity $\epsilon \equiv p/T$ is a comoving
invariant.  The differential number density for small lepton number in
this case is then
\begin{equation}
dn_{\nu_\alpha} \approx
\left({\frac{n_\gamma}{4 \zeta\left(3\right)}}\right)
{\frac{\epsilon^2 d\epsilon}{e^{\epsilon}+1}}.
\end{equation}

\subsection{Quantitative formulation for the early universe}

If one makes the assumption that the only net lepton number in the
universe is that required for electric charge neutrality ({\it i.e.},
half of the baryon number when there are equal numbers of protons and
neutrons), then the finite density potential usually remains
negligible.  In some cases where $\delta m^2 < 0$, an initially small
asymmetry like this can be amplified by matter-enhanced active-sterile
conversion \cite{fvamp,shi96}.

The finite density potential in the early universe could be dominated
by asymmetries in the lepton number, and so is often referred to as
the ``lepton potential.''  It takes the form
\begin{equation}
\label{vl}
V^D = \frac{2 \sqrt{2} \zeta (3)}{\pi^2}\,G_{\rm F} T^3 \left({\cal
L}^\alpha \pm \frac{\eta}{4}\right),
\end{equation}
where we take \lq\lq $+$\rq\rq\ for $\alpha=e$ and \lq\lq$-$\rq\rq\ for
$\alpha=\mu,\tau$.
Here we define the net driving lepton number ${\cal{L}^\alpha}$ in terms
of the lepton numbers in each active neutrino species as
\begin{equation}
\label{L}
{\cal{L}^\alpha} \equiv 2 L_{\nu_\alpha} +
\sum_{\beta\neq\alpha}{L_{\nu_\beta}}
\end{equation}
with the final sum over the active neutrino flavors other than
$\nu_\alpha$. Note that in Eq.\ (\ref{vl}), the baryon-to-photon ratio
is $\eta$, not to be confused with neutrino degeneracy parameter.

The total weak potential, $V(p,T)=V^D(T)+V^T(p,T)$,
experienced by an active neutrino $\nu_\alpha$ is approximately
\begin{equation}
\label{pot}
V\left(p,T\right) \approx \left(40.2\,{\rm
eV}\right){\left( {\frac{{\cal{L}}}{{10}^{-2}}}\right)}
{\left({\frac{T}{\rm GeV}}\right)}^3 - B p {\left({\frac{T}{\rm
GeV}}\right)}^4.
\end{equation}
As noted above, the thermal potential $V^T$ must take into account the
presence of populated leptons of the same flavor.  For $\nu_e$ this is
required at all temperatures where the neutrinos are coupled; for
$\nu_\mu$, the thermal muon term should be included at temperatures $T
\gtrsim 20 {\rm\,MeV}$, where the $\mu$ is populated; and for
$\nu_\tau$, the thermal term should be included at $T \gtrsim
180{\rm\,MeV}$. Therefore, the coefficient $B$ takes on the values:
\begin{equation}
\label{B}
B \approx \begin{cases}
10.79\,{\rm eV} &
\alpha=e
\\
\ 3.02\,{\rm eV} &
\alpha=\mu,\tau
\end{cases}
\end{equation}
for $T \lesssim 20\rm\, MeV$;
\begin{equation}
B \approx \begin{cases}
10.79\,{\rm eV} &
\alpha=e,\mu
\\
\ 3.02\,{\rm eV} &
\alpha=\tau
\end{cases}
\end{equation}
for $20\,{\rm MeV} \lesssim T \lesssim 180\rm\, MeV$;
\begin{equation}
B \approx 10.79\,{\rm eV} \quad \alpha=e,\mu,\tau
\end{equation}
for $T \gtrsim 180\rm\, MeV$.

\begin{figure}
\includegraphics[width=3.3truein]{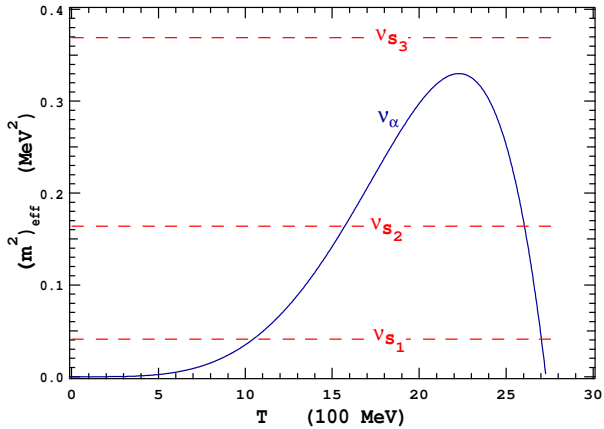}
\caption
{\small An example of the temperature evolution of the
active and sterile neutrino effective mass-squared, $m^2_{\rm
eff}$.  The active neutrino $m^2_{\rm eff}$ is dominated by a positive
finite density potential at lower temperatures and turns
over when the thermal potential dominates.  Resonance occurs at level
crossings, where the active and sterile $m^2_{\rm eff}$ tracks
intersect.  }
\label{potfig}
\end{figure}

The trend of the weak potential experienced by an active neutrino
species $\nu_\alpha$ is clear. When the quantity ${\cal{L}}\pm\eta/4$
is sufficiently large and positive (negative for $\bar\nu_\alpha$) the
potential will rise with increasing temperature, reach a maximum, and
then turn over and eventually become negative at a high temperature
where thermal terms dominate. The potential as a function of
temperature for some representative parameters is shown as the solid
line in Fig.~\ref{potfig}. With this behavior it is obvious that
neutrino mass level crossings in the temperature regime {\it not}
dominated by the thermal terms are possible only if the vacuum mass
eigenvalues most closely associated with some sterile neutrinos are
larger than those most closely associated with $\nu_\alpha$.

\section{Non-equilibrium Production: The Sterile Neutrino Boltzmann Equation}
\label{boltzsect}

The Boltzmann equation gives the evolution of the phase-space density
distribution function $f(p,t)$ for a particle species.  For sterile
neutrinos in the early universe, it can be written as
\cite{kolbturner,bernstein}
\begin{align}
&\frac{\partial}{\partial{t}}f_s(p,t) - H\,p\,
\frac{\partial}{\partial{p}}f_s(p,t) = \nonumber\\
&\quad \sum_i
\int{\Gamma_i(p_{\alpha}^\prime,p)f_{\alpha} 
(p^\prime_\alpha,t)\left[1-f_s(p,t)\right]
d^3 p^\prime_\alpha} \nonumber\\
&\qquad -\int{\Gamma_i(p_{\alpha}^\prime,p)
f_s(p,t) \left[1-f_{\alpha}(p^\prime_\alpha,t)\right] d^3 p^\prime_\alpha},
\label{classicalboltz}
\end{align}
where the rate of scattering production of $\nu_s$ corresponding to a
particular channel $i$ [see Eq.\ (\ref{interact})] is $\Gamma_i$.  We
make two approximations regarding the scattering kernels
$\Gamma_i(p^\prime,p)$: (1) they are isotropic; and (2) they are
conservative.

The first of these approximations is completely justified given a
homogeneous and isotropic universe.  The second---that an active
neutrino scatters into a sterile state of the same energy---is made
purely to ease the computational complexity of following the evolving
system of neutrinos.  We note, however, that Ref.\ \cite{dh} estimates
the effects of relaxing this approximation in the context of a
semi-analytic calculation and find that it makes little qualitative or
quantitative difference in the results.  

Certainly, however, a better treatment could be extended to the
scattering kernel.  This could, for example, change the conditions
required for coherence.  For example, we note that the effects
described in Ref.\ \cite{dolgovnonres} may be important, especially
regarding where coherence breaks down.  We show, however, in Sec.\
\ref{leptressect} that coherent production of sterile neutrinos at MSW
resonances is not important in our favored parameter regime.

In our discussion, we take $f_i(p,t)$ to be the momentum- and
time-dependent distribution function of active ($\alpha=e,\mu,\tau$)
and sterile ($s$) neutrinos.  The distribution functions $f_\alpha$
are given by Eq.\ (\ref{distfunc}). The second left hand side term
arises from the redshifting of the distribution.

However, Eq.\ (\ref{classicalboltz}) is semi-classical in its
evolution of the neutrino distributions.  To exactly follow the full
quantum development, including such quantum effects as damping and
off-diagonal contributions to the neutrino Hamiltonian in matter
\cite{qf95}, one must follow the time evolution of the density matrix
\cite{dolgov81,harstod81,siglraffelt,mckellarthomson}. One can
approximate the effects of quantum damping through a damped conversion
rate $\Gamma(\nu_\alpha\rightarrow\nu_s;p,t)$.  Fermi blocking effects
are taken to be negligible, leaving the Boltzmann equation as
\begin{align}
&\frac{\partial}{\partial{t}}f_s(p,t) - H\,p\,
\frac{\partial}{\partial{p}}f_s(p,t)
\nonumber\\ 
&\qquad\approx\Gamma(\nu_\alpha\rightarrow\nu_s;p,t)
\left[f_\alpha(p,t)-f_s(p,t)\right].
\label{boltzmann}
\end{align}
The reverse term is important for the $L\neq 0$ case, where $f_\alpha$
and $f_s$ can become comparable for certain regions of the neutrino
momentum distributions.

The conversion rate to sterile neutrinos is just the product of half
of the {\it total} interaction rate, $\Gamma_\alpha$, of the neutrinos
with the plasma and the probability that an active neutrino has
transformed to a sterile:
\begin{equation}
\Gamma(\nu_\alpha\rightarrow \nu_s; p,t) \approx
\frac{\Gamma_\alpha}{2} \langle P_m(\nu_\alpha \rightarrow
\nu_s;p,t)\rangle.
\label{conversrate}
\end{equation}
This probability $P_m$ depends on the amplitude of the matter mixing
angle and the quantum damping rate $D(p)=\Gamma_\alpha(p)/2$ $[\bar
D(p)=\bar\Gamma_\alpha(p)/2]$ for neutrinos [antineutrinos]
\cite{stodolsky,earlyfw,dll,volkaswong},
\begin{widetext}
\begin{eqnarray}
\langle P_m(\nu_\alpha\rightarrow\nu_s; p,t) \rangle &\approx& \frac{1}{2}
\frac{\Delta(p)^2 \sin^2
2\theta}{\Delta(p)^2\sin^2 2\theta + D^2(p) +
[\Delta(p) \cos 2\theta - V^D - V^T(p)]^2} 
\label{avgprob}\\
\langle P_m(\bar{\nu}_\alpha\rightarrow\bar{\nu}_s; p,t) \rangle &\approx&
\frac{1}{2} \frac{\Delta(p)^2 \sin^2 2\theta}{\Delta(p)^2\sin^2
2\theta + \bar{D}^2(p) + [\Delta(p) \cos 2\theta + V^D - V^T(p)]^2}
\label{avgprobnubar}
\end{eqnarray}

The full Boltzmann equation then is
\begin{eqnarray}
\frac{\partial}{\partial{t}}f_s(p,t) - H\, p\,
\frac{\partial}{\partial{p}}f_s(p,t) &\approx&
\frac{\Gamma_\alpha(p)}{2} \langle P_m(\nu_\alpha
\rightarrow \nu_s, t_{\rm in} + \tau)\rangle_\tau
\left[f_\alpha(p,t) - f_s(p,t)\right] \nonumber\\ 
&\approx& \frac{\Gamma_\alpha(p)}{2} \sin^2 2\theta_m
\left[1+\left(\frac{\Gamma_\alpha(p) l_m}{2}\right)^2\right]^{-1}
\left[f_\alpha(p,t) - f_s(p,t)\right] \label{fullboltz}\\ 
&\approx& \frac{1}{4}
\frac{\Gamma_\alpha(p) \Delta^2 (p) \sin^2 2\theta}{\Delta^2 (p) \sin^2
2\theta + D^2(p) + \left[\Delta (p) \cos 2\theta - V^L -
V^T(p)\right]^2} \left[f_\alpha(p,t) -
f_s(p,t)\right],
\nonumber
\end{eqnarray}
\end{widetext}
where $\{1+[\Gamma_\alpha(p) l_m/2]^2\}^{-1}$ is the
damping factor.  There are analogous equations for Eqs.\
(\ref{classicalboltz})-(\ref{conversrate}) and Eq.\ (\ref{fullboltz})
for antineutrinos.

Previous calculations have approached the solution of this equation
for the case of negligible lepton number ($L\approx 0$) analytically
and have been restricted to the short epoch just prior to BBN where
the Hubble expansion rate $H$ (evolution of the scale factor) and
time-temperature relations are simple power laws (the temperature is
proportional to the inverse scale factor) \cite{dw,dh}.  Such
approximations allow the reduction of the left-hand side of Eq.\
(\ref{fullboltz}) to a single term.  With the approximations that
lepton number ${\cal L}$ is always negligible, that the thermal term
$V^T$ is not modified by population of leptons, that the interaction
rate is not enhanced due to population of scatterers, that quantum
damping is never important, and that the reverse rates
($\nu_s\rightarrow\nu_\alpha$) are always negligible, then the right
hand side can also be considerably simplified, and the solution for
the sterile neutrino dark matter abundance is reduced to a simple
integral. It is obvious, however, that many if not all of these
approximations are eventually invalid over at least some of the
parameter range of interest for sterile neutrino dark matter.

For example, in order to probe cases where sterile neutrino dark
matter production may lie above the QCD transition ($T \gtrsim
100\rm\, MeV$), and in order to improve the accuracy of the predicted
dark matter contribution, we must extend our calculation to epochs
where the scale factor-temperature and time-temperature relations are
not simple power laws (see the Appendix for details). The interactions
contributing to $\Gamma_\alpha(p)$ which produce (and remove) sterile
neutrinos are
\begin{eqnarray}
\nu_\alpha + \nu_\beta &\rightleftharpoons& \nu_\alpha + \nu_\beta \nonumber\\
\nu_\alpha + l^\pm &\rightleftharpoons& \nu_\alpha + l^\pm \\
\nu_\alpha + q\;\; &\rightleftharpoons& \nu_\alpha + q \nonumber\\
\nu_\alpha + \nu_\alpha &\rightarrow& l^+ + l^-  \nonumber .
\label{interact}
\end{eqnarray}
Here, the $\nu_\alpha$ represent either neutrinos or anti-neutrinos,
as appropriate, and $q$ and $l$ are any populated quark and charged
lepton flavors.  The total interaction rate at temperatures $1 {\rm\,
MeV} \lesssim T \lesssim 20 {\rm\, MeV}$ due to interactions of the
neutrinos among themselves and the $e^\pm$ pairs is
\begin{equation}
\Gamma_\alpha (p) \approx \begin{cases}1.27\, G_{\rm F}^2 p T^4, \qquad
\alpha=e, \\ 0.92\, G_{\rm F}^2 p T^4,\qquad \alpha=\mu,\tau .\end{cases}
\end{equation}
At higher temperatures, other leptons and quarks are populated and
contribute to the neutrino interaction rate.  In our calculations, we
have included the enhancement of the interaction rate due to the
presence of these new particles in the plasma at high temperatures.
In particular, a significant increase to the scattering rate results
at temperatures above the QCD scale.  Interestingly, the results
of the production of sterile neutrino dark matter therefore depend on
the temperature of the QCD transition, where the quark-antiquark pairs
annihilate and are incorporated into color singlets.

The high scattering rate characteristic of the environment of the
early universe not only serves to populate the sterile neutrino sea,
but can also suppress the production of sterile neutrinos when the
matter oscillation length is large compared to the mean free path of
the neutrinos.  When the oscillation length is much larger than the
mean free path, the probability that an active neutrino has
transformed into a sterile state becomes very small.  It can be shown
that such scattering can force a quantum system to not evolve from the
initial state \cite{qze,stodolsky}.  Essentially, each scattering
event resets the phase of the developing neutrino state $\vert
\Psi_{\nu_\alpha}
\rangle$. This is the so-called quantum Zeno effect. 

Consider the right hand side collision term in Eq.\
(\ref{boltzmann}). The sterile neutrinos are initially not in thermal
equilibrium, and therefore the reverse processes are initially
unimportant.  They will, however, become more important as the sterile
neutrino sea is populated. Because the sterile neutrinos are {\it
never} in equilibrium, the usual simplifying principle of steady-state
equilibrium \cite{kolbturner,gongel} cannot be made.  Therefore, to
calculate the production of sterile species, we must start with the
Boltzmann equation in its ``unintegrated'' form Eq.\
(\ref{fullboltz}).

We have directly solved the Boltzmann equation numerically for the
$L\approx 0$ case ({\it e.g.}, $L\approx 10^{-10}$), and found the
amount of sterile neutrino dark matter produced for a broad range of
sterile neutrino mass and mixing angles with each of the three active
neutrino flavors.  In treating the redshift of the sterile
distribution, we have greatly accelerated the numerical calculation by
following the redshift of the sterile neutrino distribution function
$f_s$ by redshifting momenta as $p\propto t^{-1/2}$, which is always
true in a radiation dominated universe (a necessary condition for BBN
and the CMB).  This numerical method allows us to correctly follow the
redshifting of the sterile neutrino without the second term of the
Liouville operator in the Boltzmann equation (\ref{fullboltz}), while
active neutrinos may be reheated via annihilation of other species.

\begin{figure}
\vskip 3mm
\includegraphics[width=3.3truein]{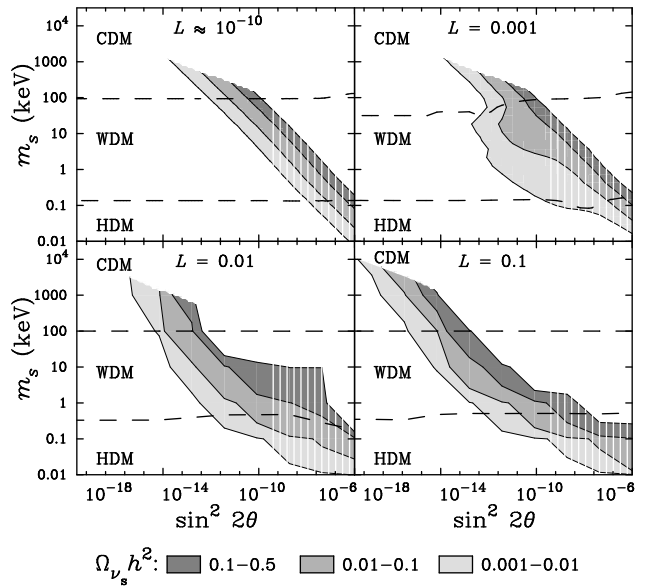}
\caption{\small Regions of $\Omega_{\nu_s} h^2$ produced by resonant
and nonresonant $\nu_e \leftrightarrow \nu_s$ neutrino conversions for
selected net lepton number $L$, after applying all constraints (see
Secs.\ \ref{cosmoconstraints} and \ref{supernovaconstraints}). Regions
of parameter space disfavored by supernova core collapse
considerations are shown with vertical stripes.}
\label{omegaepanel}
\end{figure}

\begin{figure}
\vskip 3mm
\includegraphics[width=3.3truein]{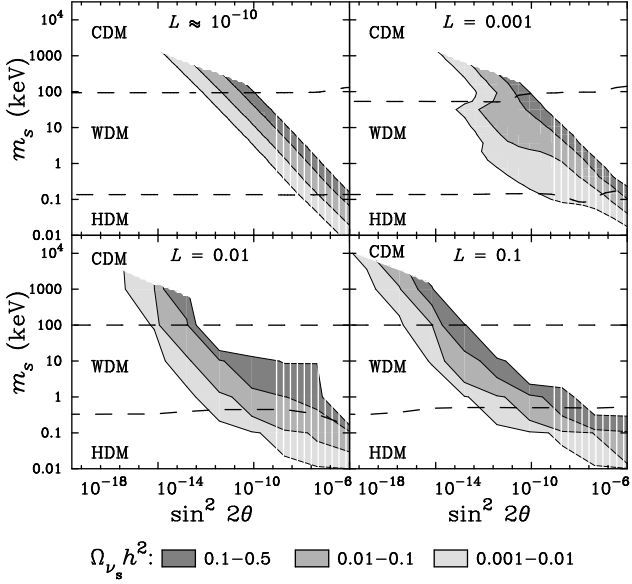}
\caption
{\small Same as Fig.\ \ref{omegaepanel}, but for $\nu_\tau
\leftrightarrow \nu_s$.}
\label{omegataupanel}
\end{figure}

The contribution to the closure fraction of the universe, meeting all
constraints (see Secs.\ \ref{cosmoconstraints} and
\ref{supernovaconstraints}) is shown in Fig.\ \ref{omegaepanel} for
$\nu_e\rightleftharpoons\nu_s$ and Fig.\ \ref{omegataupanel} for
$\nu_\tau\rightleftharpoons\nu_s$.  A general fit to our results for
nonresonant production is
\begin{equation}
\Omega_{\nu_s} h^2 \approx 0.3 \left(\frac{\sin^2
2\theta}{10^{-10}}\right) \left({\frac{m_s}{100\,\rm keV}}\right)^2.
\end{equation}
The maximum rate of sterile neutrino production occurs at temperature
\cite{barbk,dw}
\begin{equation}
T_{\rm max} \approx 133{\rm\,
MeV}\left(\frac{m_s}{1{\rm\,keV}}\right)^{1/3}.
\label{tmax}
\end{equation}

The closure fraction contribution for $\nu_\mu\rightleftharpoons\nu_s$
is nearly identical to $\nu_\tau\rightleftharpoons\nu_s$.  The
particular flavor(s) of the active neutrino with which the sterile
neutrino mixes does determine to some extent the ultimate closure
fraction.  However, this flavor dependence is negligible for larger
$\nu_s$ masses since, from Eq.\ (\ref{tmax}), for sterile neutrinos
with $m_s \gtrsim 2\rm\, keV$, $T_{\rm max} \gtrsim 180\rm\, MeV$. At
these temperatures, $e^\pm$, $\mu^\pm$ and even $\tau^\pm$ (due to the
high entropy of the universe) are populated significantly, so that the
thermal potentials are identical for all flavors.  Therefore, the mass
fractions produced for all flavors with sterile neutrino masses $m_s
\gtrsim 2\,\rm keV$ are very similar.  

The similarity of the results of production for $\nu_e$, $\nu_\mu$,
and $\nu_\tau$ mixing with a sterile neutrino is at odds with the
calculation of Ref.\ \cite{dh}.  There are several differences between
our treatment and that of Ref.\ \cite{dh} that can account for the
disparity in results.  First, the primary cause for discrepancy for
the $\nu_\tau\rightleftharpoons \nu_s$ neutrino mixing case is likely
to lie in the fact that $\tau$ leptons are significantly populated at
temperatures one-tenth of their mass ($T \gtrsim 177\rm\, MeV$) due to
the high-entropy of the universe.  This modifies the thermal term in
Eq.\ (\ref{vt}) in an important way, not included in Ref.\
\cite{dh}. In our work, we follow the number and energy density of tau
leptons explicitly in our numerical evolution.  One can see from Eq.\
(\ref{tmax}) that for masses of sterile neutrinos greater than about
$2\rm\, keV$, production occurs at a temperature where the $\tau$
lepton is significantly populated. Second, the population of massive
species of scatterers ($\mu$ and $\tau$ leptons and $u,d,s,c,b$
quarks) which enhance the scattering rate is not included in Ref.\
\cite{dh}, but are included in our calculations. Third, effects of
re-heating on the dilution of sterile neutrino dark matter are treated
only approximately in Ref.\ \cite{dh}.  We explicitly include
reheating and dilution in our calculation through our treatment of the
time-temperature evolution of active and sterile neutrinos, as
described in the Appendix.


\section{Initial Lepton Number and Resonant Production}
\label{leptressect}

In this section, we examine the resonant production of sterile
neutrino dark matter for a range of initial lepton numbers.  We will
extend the treatment of this issue given in Shi and Fuller
\cite{cooldm} considering the effects
of particle annihilation and reheating and enhanced scattering rates on
coherent and incoherent evolution through MSW resonances.  
This allows us to consider resonant $\nu_\alpha\rightleftharpoons
\nu_s$ conversion at high temperature epochs in the early universe, up
to the electroweak transition regime at $T\approx 100\rm\, GeV$.

Resonance, or mass level crossings as seen in Fig.\ \ref{potfig}, are
characterized by maximal effective matter mixing angles
$(\theta_m)_{\rm res} = \pi/4$, where the $\nu_\alpha \rightarrow
\nu_s$ conversion rate is, consequently, enhanced. 
The resonance condition is
\begin{align}
& \Delta (p) \cos 2\theta - V^L - V^T(p) = 0\nonumber \\
& \Delta (p) \cos
2\theta - \left(40.2\,{\rm eV}\right){\left\{
{\frac{{\cal{L}}\pm\eta/4}{{10}^{-2}}}\right\}} {\left({\frac{T}{\rm
GeV}}\right)}^3 \nonumber\\
&\quad + B \epsilon {\left({\frac{T}{\rm
GeV}}\right)}^5
 = 0\nonumber\\
\label{rescondition}
&\left(\frac{m_{s}}{1\,{\rm keV}}\right)^2 \cos 2\theta \approx
8.03\ \epsilon\ {\left\{
{\frac{{\cal{L}}\pm\eta/4}{{10}^{-2}}}\right\}}
{\left({\frac{T}{100\,\rm MeV}}\right)}^4 \nonumber\\
&\qquad\qquad + 2 \epsilon^2
\left(\frac{B}{\rm keV}\right) {\left({\frac{T}{100\,\rm
MeV}}\right)}^6.
\end{align}

Choose one of the horizontal dotted lines laying below the peak of the
solid line in Fig.\ \ref{potfig}. This indicates the \lq\lq mass
track\rq\rq\ for a sterile neutrino species. This is, of course,
simply the vacuum mass-squared value ($m^2_{\rm eff})_s =
m^2_s$, and is flat and independent of temperature. 
The effective mass-squared track for an active neutrino $\nu_\alpha$
could be as shown in Fig.\ \ref{potfig}. Mass level crossings
(resonances) in the $\nu_\alpha\rightleftharpoons \nu_s$ system can
occur when $m^2_s$ lies below the peak value of $(m^2_{\rm
eff})_{\nu_\alpha}$.  This peak will occur at temperature
\begin{align}
\label{Tmsmax}
T_{\rm PEAK} &\approx
{\left({\frac{2}{3}}\right)}^{1/2} T_0 \nonumber \\ 
&\approx
{\left({\frac{4\sqrt{2}\zeta(3)}{3 \pi^2 }}\right)}^{1/2} 
{\epsilon}^{-1/2} {\left\{ {\cal{L}}
\pm\eta/4\right\}}^{1/2} \nonumber\\
&\quad\times\left[\frac{G_F ({\rm GeV})^5}{B}\right]^{1/2} 
\nonumber \\ 
&\approx (2.98{\rm\, GeV}) {\epsilon}^{-1/2}
{\left({\frac{\cal{L}}{{10}^{-2}}}\right)}^{1/2} ,
\end{align}
where $T_0$ is the high temperature at which the effective
mass-squared track crosses zero, and where in the last numerical
expression we have assumed that $T>180\,\rm MeV$.

Note that the peak value of effective mass-squared, or equivalently, the
largest value of $\delta m^2 \cos2\theta$ for which a level crossing can
occur is,
\begin{align}
\label{delmax}
m^2_{\rm eff}\left(T_{\rm PEAK}\right) &= {\left(
\delta m^2 \cos2\theta \right)}_{m^2_{\rm eff(PEAK)}}\\
&\approx {\left(
{\frac{4\sqrt{2}\zeta(3)}{ 3 \pi^2 }}\right)}^{3} {\frac{{\left\{
{\cal{L}}\pm\eta/4\right\}}^{3}}{ 
\epsilon}}\left[\frac{G_F^3 ({\rm GeV})^{10}}{B^2}\right] \nonumber.
\end{align}
For example, we can show that the largest $\nu_s$ mass which can have
a resonance with a $\nu_e$ is roughly 
\begin{equation}
(m_s)_{\rm PEAK} \approx \frac{406 {\rm\, keV}}{\epsilon^{1/2}}
{\left({\frac{\cal{L}}{{10}^{-2}}}\right)}^{3/2}.
\end{equation}

\begin{figure}
\includegraphics[width=3.3truein]{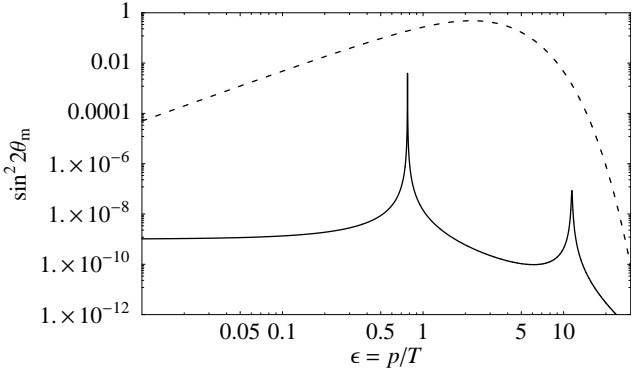}
\caption
{\small The effective matter mixing $\sin^2 2\theta_m$ is shown for a
case with two resonances (solid line).  For this case, $m_s = 1\,\rm
keV$, $\sin^2 2\theta = 10^{-9}$, $L = 6 \times 10^{-4}$, and $T =
130\,\rm MeV$. The dashed line is the active neutrino energy
distribution.  }
\label{doubleres}
\end{figure}

As the universe expands, an active neutrino species $\nu_\alpha$ will
encounter two resonances with a sterile neutrino species, as long as
this sterile species has $m^2_s$ less than the peak value of $m^2_{\rm
eff}$ in Eq.\ (\ref{delmax}). At some epochs, both resonances may be
present in a given active neutrino spectrum since the resonance
condition, Eq.\ (\ref{rescondition}), has more than one zero.  This
behavior is shown for a particular case in Fig.\ \ref{doubleres}.

What is the effect of resonance on sterile neutrino production in the
early universe?  The answer to this question depends on the rate of
incoherent production at the resonance, whether the neutrinos are
coherent as they pass through resonance and, if they are coherent,
the degree of adiabaticity characterizing the evolution of
the neutrino flavor amplitudes through the resonance.  There are two
processes by which resonance affects neutrino conversion in the early
universe: (1) enhanced incoherent conversion of those neutrinos at the
resonance and (2) coherent MSW transformation of neutrinos passing
through the resonance.  The rate of incoherent production can be
calculated numerically through the semi-classical Boltzmann evolution, Eq.\
(\ref{fullboltz}).

The statistical formulation of both coherent and incoherent production
can be described by the time evolution of the density matrix for the
two neutrino states
\cite{dolgov81,harstod81,siglraffelt,stodolsky,mckellarthomson,qf95}.
We show below that the masses and mixing angles of interest for
sterile neutrino dark matter considered here give coherence across the
relevant resonance width, but usually imply that the neutrino
amplitude evolution through resonance is nonadiabatic. (Adiabatic
evolution at lower mixing angle could occur at epochs where the
entropy is being transferred from annihilating particles and where, as
a consequence, the temperature and density do not change rapidly or
are constant with time, {\it e.g.}, the QCD transition \cite{afprep}.)

Though the expansion rate of the universe scales as $T^2$, the active
neutrino scattering rate scales as $G_{\rm F}^2 T^5$. For neutrino flavor
evolution through resonances, the coherence condition will be met only
when the inverse of the scattering rate (the active neutrino mean free
path) is much larger than the resonance width or
\begin{eqnarray}
\label{coh}
T&\ll& {\left({\frac{4\pi^3}{5}}\right)}^{1/6}{\frac{g^{1/6}}{m_{\rm
pl}^{1/3} G_{\rm F}^{2/3} \tan^{1/3}2\theta}}\nonumber\\ 
&\approx& {140\,{\rm MeV}
\frac{{\left({g/100}\right)}^{1/6}}
{{\left(\sin^2 2\theta/{10}^{-10}\right)}^{1/6}}}.
\end{eqnarray}
(Here, $m_{\rm pl}\approx 1.22\times 10^{22}\rm\, MeV$ is the Planck
mass.)  Therefore, for the mixing angles relevant here ($\sin^22\theta
\lesssim 10^{-10}$), at temperatures below $140\,\rm MeV$, the
resonance could in principle drive coherent $\nu_\alpha
\rightleftharpoons \nu_s$ transformation.  The efficiency of MSW
conversion relies on the adiabaticity of the evolution through the
resonance region.

The width of the resonance is the product of the local density scale
height of weak charges and $\tan2\theta$. In turn, the biggest share
of the density scale height is determined by the expansion rate of the
universe, $H\approx (8 \pi^3/90)^{1/2} g^{1/2} T^2/m_{\rm pl}$, where
the statistical weight in relativistic particles, $g$, has
contributions from both bosons $g_b$ and fermions $g_f$:
\begin{equation}
g=\sum_i{(g_b)_i}+7/8\sum_i{(g_f)_i}. 
\label{gstar}
\end{equation}
In fact, once
neutrino flavor transformation begins, the inherent nonlinearity of
this process can have a sizable, even dominant effect on the density
scale height. Ignoring this, the resonance width expressed in time is
$\delta t \approx (2/3) t (\tan2\theta) = (1/3) H^{-1}\tan2\theta$,
where $t\approx (1/2) H^{-1}$ is the age of the universe at an epoch
with temperature $T$ in radiation dominated conditions. [The particle
horizon is $H^{-1}$, so that, absent large scale neutrino flavor
transformation, the resonance width is a constant fraction $(1/3)
\tan2\theta$ of the horizon scale.]

The effective matter mixing angle for the oscillation channel
$\nu_\alpha\rightleftharpoons\nu_s$ at an epoch with temperature $T$
is
\begin{equation}
\label{thetaeff}
\sin^22\theta_m = {\left\{ 1+ {\frac{{\left[ 1-2 \epsilon T
V/\left(\delta
m^2\cos2\theta\right)\right]}^2}
{\tan^22\theta}}
\right\}}^{-1}.
\end{equation}
At resonance, $\sin^22\theta_m =1$, so that one resonance width off
resonance this effective mixing will have fallen to $\sin^22\theta_m
=1/2$. Clearly, the change in effective weak potential over this
interval is $\delta V \approx \delta m^2 \sin2\theta/(2\epsilon T_{\rm
res})$, where $T_{\rm res}$ is the resonance temperature for neutrino
spectral parameter $\epsilon$. From this it can be seen that
${\left(\delta V/V\right)}_{\rm res} =\tan2\theta$ and it follows that
the resonance width is
\begin{equation}
\label{reswidth}
\delta t ={\frac{\delta t}{\delta V}} \delta V \approx {\Bigg
\vert {\frac{1}{V}} \cdot {\frac{dV}{dt}}\Bigg\vert}^{-1}_{\rm res}
\tan2\theta .
\end{equation}
The neutrino oscillation length at resonance is $l^{\rm res}_m=(4\pi
\epsilon T_{\rm res})/(\delta m^2 \sin2\theta)= 2\pi/\delta V$, and
the adiabaticity parameter is proportional to the ratio $\delta
t/l^{\rm res}_m$, and is defined as
\begin{equation}
\label{gamma}
\gamma \equiv 2\pi  {\frac{\delta t}{l^{\rm res}_m}}
\approx {\left( \delta V\right)}^2 
\left|\frac{d\epsilon}{dV} \Bigg/\frac{d{\epsilon}}{dt}\right|_{\rm res} .
\end{equation}
When there are many oscillation lengths within the resonance width
({\it i.e.}, when $\gamma\gg 1$) neutrino flavor evolution will be
adiabatic and we can then expect efficient flavor conversion at
resonance. [Given the small vacuum mixing angles relevant for this
work and consequent small widths, the Landau-Zener jump probability
\cite{Haxton} with the form $P_{\rm LZ} \approx \exp{\left(-\pi
\gamma/2\right)}$ gives an adequate gauge of the probability of
$\nu_\alpha \rightarrow \nu_s$ conversion at resonance, $1-P_{\rm
LZ}$.]

The physical interpretation of Eq.\ (\ref{gamma}) is straightforward.
The adiabaticity parameter is proportional to the energy width of the
resonance divided by the \lq\lq sweep rate,\rq\rq\
${d{\epsilon}}/{dt}$, of the resonance energy through the neutrino
distribution function. The resonance sweep rate is determined mostly
by the expansion rate of the universe (an inverse gravitational time
scale), but the rate of change of lepton number ${\cal{L}}$ as
$\nu_\alpha\rightarrow \nu_s $ proceeds can become important, even
paramount as lepton number is used up and ${\cal{L}} \rightarrow 0$.

The resonance energy width scaled by temperature is $\delta\epsilon
\approx \delta V {\big \vert {{d\epsilon}/{dV}} \big\vert}_{\rm res} 
\approx \epsilon_{\rm res} \tan2\theta$, where $\epsilon_{\rm res}$ 
is the resonant value of the neutrino spectral parameter at $T_{\rm
res}$. If we confine our discussion to resonances on the low
temperature side of $T_{\rm PEAK}$, where the thermal
terms in the potential can be neglected, then
\begin{eqnarray}
\label{eres}
\epsilon_{\rm res} &\approx& {\frac{\delta m^2\cos2\theta}{
{\left( 4\sqrt{2}\zeta(3)/\pi^2\right)} G_{\rm F} T^4 {\cal{L}} }}\\
&\approx& 0.1245 {\left({\frac{\delta m^2\cos2\theta}{1\,{\rm
keV}^2}}\right)} {\left({\frac{{10}^{-2}}{{\cal{L}}}}\right)}
{\left({\frac{100\,{\rm MeV}}{T}}\right)}^4 \nonumber .
\end{eqnarray}
As the universe expands and cools with time, and for a given $\delta
m^2$, the resonance will sweep through the $\nu_\alpha$ energy
distribution function from low to high neutrino spectral parameter
$\epsilon$. In this same limit of resonance below $T_{\rm PEAK}$,
the sweep rate is
\begin{equation}
\label{dedt}
{\frac{d\epsilon}{dt}} \approx 4 \epsilon H {\left(
1-{\frac{\dot{\cal{L}}}{4 H {\cal{L}}}}\right)} ,
\end{equation}
where ${\dot{\cal{L}}}$ is the time rate of change of the lepton number
resulting from neutrino flavor conversion.
Since the expansion rate scales as $H\sim T^2$, the prospects for
adiabaticity are better at lower temperatures and later epochs in the
early universe, all other parameters being the same.

From Eqs.\ (\ref{gamma}), (\ref{eres}), and (\ref{dedt}), we can
estimate that at resonance the degree of adiabaticity is 
\begin{align}
\label{gammares}
\gamma &\approx \frac{\delta m^2}{2\epsilon T_{\rm res}}
\frac{\sin^22\theta}{\cos 2\theta}
\left(4\frac{\dot{T}}{T} + \frac{\dot{\cal L}}{\cal L}\right)^{-1}
\\
&\approx
{\frac{3\sqrt{5}{\zeta(3)}^{3/4}}{2^{17/8}\pi^3}} {\frac{\left(\delta
m^2\right)^{1/4} m_{\rm pl}\ G_{\rm F}^{3/4} {\cal{L}}^{3/4}}{g^{1/2}
{\epsilon}^{1/4} {\Big \vert { 1-{{\dot{\cal{L}}}/{4 H
{\cal{L}}}}}\Big\vert}}}
{\frac{\sin^22\theta}{\cos^{7/4}2\theta}} \nonumber \\
&\approx {\left({\frac{m_{s}}{1\,{\rm keV}}}\right)}^{1/2}
{\left({\frac{10.75}{g}}\right)}^{1/2}
{\left({\frac{{\cal{L}}}{{10}^{-2}}}\right)}^{3/4} {\frac{1}{{\Big
\vert { 1-{{\dot{\cal{L}}}/{4 H {\cal{L}}}}}\Big\vert}}}\nonumber\\
&\quad\times\left(\frac{1}{\epsilon_{\rm res}}\right)^{1/4} {\left\{
{\frac{\sin^22\theta}{7.5\times{10}^{-10}}}\right\}},
\label{adexamp}
\end{align}
where in the second equality we assume the standard radiation-dominated
conditions and expansion rate, and where in the final equality we have
employed the approximation $\delta m^2 \approx m_{s}^2$, valid when
$m_{s}\gg m_{\nu_\alpha}$, and where we have assumed that the vacuum
mixing angle is small.  Here we see that for the mixing angles allowed
by our constraints, $\sin^2 2\theta < 10^{-9} (3\times 10^{-10})$ for
$\nu_\mu,\nu_\tau (\nu_e)$ mixing with sterile neutrinos, and masses
$m_s \gtrsim 1\rm\, keV$, the resonance is not adiabatic.

We conclude that the main effect of resonance is enhancement of
scattering-induced incoherent conversion of neutrinos with energies in
the resonant region.  Therefore, the formulation of the semi-classical
Boltzmann Equation (\ref{fullboltz}) is appropriate for calculating the
total production of sterile neutrinos in the early universe.

\begin{figure}
\includegraphics[width=3.3truein]{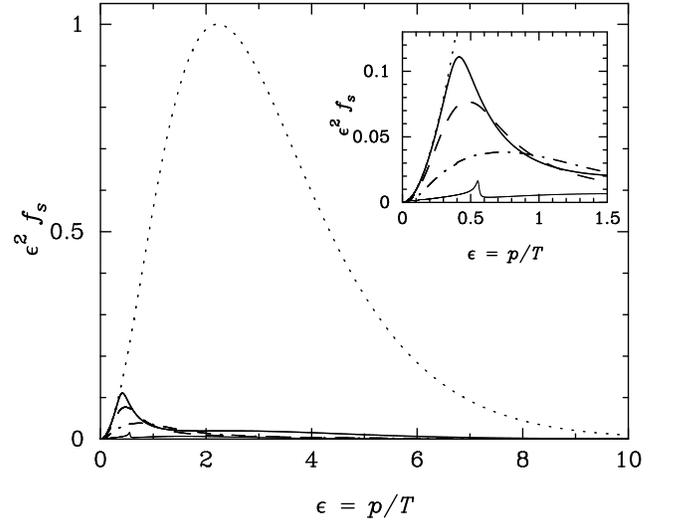}
\caption
{\small The sterile neutrino distribution for four cases of resonant
and non-resonant $\nu_e \leftrightarrow \nu_s$,
as described in the text.  The dotted line is a normalized active
neutrino spectrum.  The thick-solid, dashed, dot-dashed, and
thin-solid lines correspond to cases (1)--(4), respectively.  The inset
shows a magnified view of the low momenta range of the distributions.}
\label{spectra}
\end{figure}

The results of our numerical calculations can be seen in Fig.\
\ref{omegaepanel} for $\nu_e\rightleftharpoons\nu_s$ and in Fig.\
\ref{omegataupanel} for $\nu_\tau\rightleftharpoons\nu_s$ for the
cases where initially $L = 0.001,0.01,0.1$.  The calculation includes
both nonresonant scattering production and matter-enhanced (resonant)
production.  Examples of the resulting sterile neutrino energy spectra
are shown in Fig.\ \ref{spectra}. Resonantly produced sterile
neutrinos tend to have energy spectra appreciably populated only at
the low $\epsilon$ end.  This results from the resonant energy
starting at the lowest momenta and moving through higher momenta
neutrinos [see Eq.\ (\ref{eres})] as the universe cools and lepton
number is depleted through conversion into a sterile neutrino population.

Figure\ \ref{spectra} shows the resulting spectrum for four sample
cases of sterile neutrino dark matter production:
\begin{enumerate}[(1)]
\item $m_s = 0.8$ keV, $\sin^2 2\theta = 10^{-6}$, $L_{\rm init} =
0.01$, resulting in $\Omega_{\nu_s} h^2 = 0.25$ and $\langle p /T\rangle =
2.9$;
\item $m_s = 1$ keV, $\sin^2 2\theta = 10^{-7}$,
$L_{\rm init} = 0.01$, resulting in $\Omega_{\nu_s} h^2 = 0.13$ and $\langle
p/T \rangle= 1.8$;
\item $m_s = 1$ keV, $\sin^2
2\theta = 10^{-8}$, $L_{\rm init} = 0.01$, resulting in $\Omega_{\nu_s} h^2
= 0.10$ and $\langle p/T\rangle = 2.0$;
\item $m_s = 10$ keV, $\sin^2 2\theta = 10^{-8}$, $L_{\rm init} = 0.001$,
resulting in $\Omega_{\nu_s} h^2 = 0.57$ and $\langle p/T\rangle = 2.3$.
\end{enumerate}
A particularly interesting case is (4), where the resonance passes
through the distribution during the QCD transition, where the
disappearance of degrees of freedom heats the photon and neutrino
plasma, forcing the universe to cool more slowly.  For this period,
the resonance moves much more slowly through the spectrum and is
consequently more efficient in $\nu_\alpha \rightarrow \nu_s$
conversion through that region of the neutrino energy spectra.  This
produces a ``spike'' in the sterile neutrino distribution (see
Ref.\ \cite{afprep}).

\section{Collisionless Damping Scale: Hot, Warm or Cold Neutrinos}
\label{colldampsect}

\begin{figure}
\includegraphics[width=3.3truein]{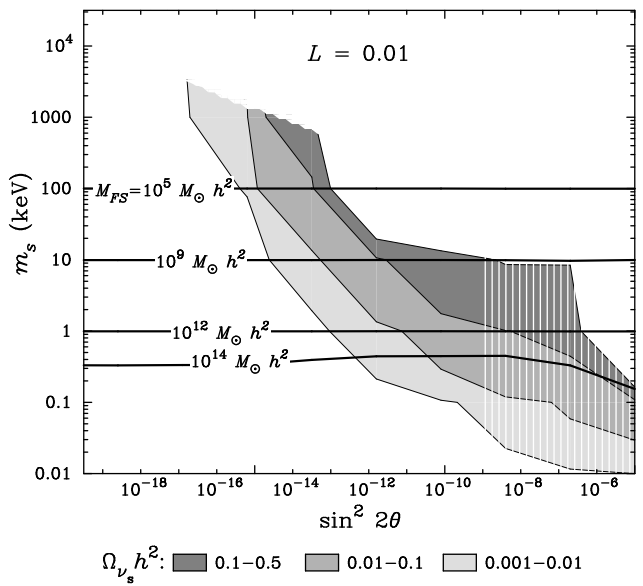}
\caption
{\small The mass within the sterile neutrino free streaming length at
matter-radiation equality ($M_{FS}$) in the
$\nu_\tau\leftrightarrow\nu_s$, $L = 0.01$ case. Regions of parameter
space disfavored by supernova core collapse considerations are shown
with vertical stripes.}
\label{mfsfig}
\end{figure}

Given the sterile neutrino energy spectrum and number density, the
transfer function for dark matter models can be calculated.  Here, we
instead give a rough guide based on the free streaming length at
matter-radiation equality, $\lambda_{FS}$.  Structures smaller than
$\lambda_{FS}$ are damped.  This is because at early epochs where
sterile or active neutrino species are relativistic, they can freely
flow out of the regions of sizes smaller than $\lambda_{FS}$ (very
roughly the horizon size at the epoch where the neutrinos revert to
nonrelativistic kinematics).  Previous numerical work has shown that
the free streaming scale is approximately \cite{kolbturner,bes}
\begin{equation}
\lambda_{FS} \approx 40\,{\rm Mpc}\left(\frac{30\rm\, eV}{m_\nu}\right)
\left(\frac{\langle p/T\rangle}{3.15}\right).
\label{lambdafs}
\end{equation}
The mass contained within the free streaming length is then
\begin{equation}
M_{FS} \approx 2.6\times 10^{11}\, M_\odot\;\left(\Omega_{m}
h^2\right)\left(\frac{1\rm\, keV}{m_\nu}\right)^3 \left(\frac{\langle p
/T\rangle}{3.15}\right)^3,
\label{mfs}
\end{equation}
where $\Omega_m$ is the contribution of all ``matter'' to closure.
These values can give a guide as to the collisionless damping scale of
sterile neutrino dark matter.  In Fig.\ \ref{mfsfig}, we show contours
of $M_{FS}$ for the $\nu_\tau\rightleftharpoons\nu_s$ sterile neutrino
production channel with $L=0.01$.  In Fig.\ \ref{mfsfig}, we assume
that the universe is critically closed ($\Omega_m = 1$).  Therefore,
the contours are consistent with sterile neutrinos being the major
constituent of dark matter near the dark gray regions where
$\Omega_{\nu_s} h^2 = 0.1-0.5$.

In Figs.\ \ref{omegaepanel} and \ref{omegataupanel}, we label the
parameter regions corresponding to $M_{FS}>10^{14} M_\odot$, $10^{5}
M_\odot<M_{FS}<10^{14} M_\odot$, and $M_{FS}<10^{5} M_\odot$ as HDM,
WDM, and CDM, respectively. These definitions are somewhat arbitrary
and here serve only as guides. Note that sterile neutrino mass by
itself does not completely determine the free streaming (or
collisionless damping) scale, since the inherent $\nu_s$ energy
spectrum at production also helps to determine the $\nu_s$ kinematics
at a given epoch. This latter effect is especially pronounced for
nonthermal energy spectra, and is responsible, {\it e.g.}, for the
non-flat collisionless damping mass scale lines in Fig.\ \ref{mfsfig}.

\section{Cosmological Constraints on Sterile Neutrinos}
\label{cosmoconstraints}

\subsection{Diffuse extragalactic background radiation}

The decay rate of a massive sterile neutrino with vacuum mixing angle
$\theta$ into lighter active neutrinos is \cite{barger}
\begin{eqnarray}
\Gamma_{\nu_s} &\approx& \sin^2 2\theta\ G_{\rm F}^2
\left(\frac{m_s^5}{768\pi^3}\right)
\label{decayrate}\\ &\approx& 8.7\times 10^{-31}
{\,\rm s^{-1}} \left(\frac{\sin^2 2\theta}{10^{-10}}\right)
\left(\frac{m_s}{1{\rm\, keV}}\right)^5.\nonumber
\end{eqnarray}
We have also included in our calculations the contributions to
$\Gamma_{\nu_s}$ from visible and hadronic decays estimated from the
partial decay widths of the $Z^0$ boson \cite{pdg}.  The rate of the
corresponding radiative decay branch is smaller by a factor of
$27\alpha/8\pi$ \cite{pal}:
\begin{eqnarray}
\Gamma_{\nu_s\gamma} &\approx& \sin^2 2\theta\ \alpha\ G_{\rm F}^2
\left(\frac{9 m_s^5}{2048\pi^4}\right)
\label{debrarate}\\ &\approx& 6.8\times 10^{-33}
{\,\rm s^{-1}} \left(\frac{\sin^2 2\theta}{10^{-10}}\right)
\left(\frac{m_s}{1{\rm\, keV}}\right)^5.\nonumber
\end{eqnarray}

That these sterile neutrinos can decay stems from the fact that they
are not truly ``sterile,'' but have an effective interaction strength
$\sim\!\sin^2_m 2\theta\ G_{\rm F}^2 = \sin^2 2\theta\ G_{\rm F}^2$,
where the last equality is valid in vacuum ({\it i.e.}, at late epochs
in the universe).  Obviously, if a particle is going to be a dark
matter candidate, it must have a lifetime $\tau$ at least of order the
age of the universe ($t_{\rm today} \gtrsim 10\rm\, Gyr$). Regions of
$m_s$ vs $\sin^2 2\theta$ parameter space where $\tau < t_{\rm
today}$ and not otherwise constrained are so labeled in Figs. \ref{L0}
and \ref{Ln0}. These parameters, corresponding to generally high
sterile neutrino masses, guarantee that these particles decay away
into lighter particles long before they clump in gravitational
potential wells.

The radiative decay channel for sterile neutrinos can provide a
constraint or a possible detection mode for some regions of the
$m_s-\sin^2 2\theta$ parameter space.  In particular, radiative decays
of sterile neutrinos occuring between CMB decoupling and today could
produce an appreciable flux of photons with energies comparable to
$m_s$.  In fact, there are firm upper limits on the flux from a
diffuse photon component.  For example, the total differential energy
flux per unit solid angle for a DEBRA component is
\cite{ressellturner,dicuskolbteplitz,kolbturner}
\begin{equation}
d{\cal F}/d\Omega \lesssim
(1{\rm\,MeV}/E)\rm\ cm^{-2}sr^{-1}s^{-1}.\label{debralimit}
\end{equation}

As technology has progressed this limit on the true diffuse background
has come down as distinct x-ray sources are resolved and their fluxes
are removed from the count \cite{comastri,chandra}.  At present the
Chandra X-ray Observatory threatens to resolve a considerable fraction
of the observed x-ray background (in the $0.5-8\,\rm keV$ band) into
point and extended sources, primarily active galaxies, QSO's, and
clusters.

Clearly, sterile neutrinos with radiative decay rates greater than an
inverse Hubble time ($H^{-1} \approx 3.09\times 10^{17}\, h^{-1}{\,\rm
s} \approx 9.78\, h^{-1} {\,\rm Gyr}$) will tend to contribute
background photons before the sterile neutrinos fall into potential
wells and form structure. These photons will then produce a DEBRA
contribution which must not exceed the overall limit in Eq.\
(\ref{debralimit}). Parameter regions violating this bound are labeled
DEBRA in Figs.\ \ref{L0} and \ref{Ln0}.

What about sterile neutrinos with much smaller radiative widths? These
steriles could be decaying at more recent epochs, even today.
However, if these are the dark matter (CDM or WDM), then they are not
diffuse, but are strongly clustered.  In Figs.\ \ref{L0} and
\ref{Ln0} the regions with widely spaced vertical lines correspond to
sterile neutrino mass and mixing properties that would give a DEBRA
component in excess of the limit (\ref{debralimit}), {\it if} these
steriles were distributed diffusely.  Since most of the decay photons
will be produced when the dark matter is in structure rather than
diffuse, this region does not as yet constitute a constraint.  Rather,
it serves to define parameters that could give interesting x-ray
fluxes in the gravitational potential wells of clusters of galaxies or
other structures, depending on redshift, the cosmological parameters,
and transfer functions and collisionless damping scales of the sterile
neutrinos.  Improved observations and models could lead to these
regions turning into true constraints and may result in more
stringent constraints or even lead to detection \cite{afp3}.

As an example, consider a large cluster of galaxies with dark matter
mass $M\approx 10^{15}\, M_\odot$. The sterile neutrino decay
luminosity in photons is
\begin{equation}
L\approx 7\times 10^{36} {\,\rm erg\, s^{-1}} \left(\frac{\sin^2
2\theta}{10^{-10}}\right) \left(\frac{m_s}{1{\rm\, keV}}\right)^5.
\end{equation}
The observed x-ray luminosity for such clusters in the $1-10\rm\, keV$
band is $L_{\rm cluster} \sim 10^{45}\rm\, erg\, s^{-1}$ \cite{gioia}.
Clearly, for sterile neutrinos in our dark matter parameter space with
masses of $20 - 30\rm\, keV$ and $\sin^2 2\theta\approx 10^{-11}$, the
predicted luminosities may be comparable to those observed.  The
energy spectra may be different, however, and this could be an avenue
for constraint.

\subsection{Cosmic microwave background}

Another constraint stems from the increase in energy density in
relativistic particles due to massive sterile neutrino decay prior to
cosmic microwave background (CMB) decoupling \cite{hrcmb}.  The
BOOMERanG/MAXIMA observations \cite{boommax} currently limit the
effective number of neutrinos ($N_\nu$) at decoupling to $N_\nu ({\rm
CMB}) < 13$ at the 95\% confidence level \cite{hannlesg}.
Measurements to higher multipole moments by the MAP and Planck probes
will be able to further limit the relativistic energy present at
decoupling.  MAP may constrain $N_\nu ({\rm CMB}) < 3.9$, and the
Planck mission could reach a limit $N_\nu ({\rm CMB}) < 3.05$
\cite{lopezcmb}. The increase in energy density due to sterile
neutrino decays was found by calculating the energy produced by decays
between active neutrino decoupling ($T\sim 1\rm\,MeV$) and photon
decoupling ($T\sim 0.26\rm\, eV$).  The BOOMERanG/MAXIMA constraint is
shown in Figs. \ref{L0} and \ref{Ln0} and the potentially stringent
future ``constraints'' by MAP and Planck are also shown.

\subsection{Big bang nucleosynthesis}

The energy density in the sterile neutrino sea at weak freeze-out just
prior to primordial nucleosynthesis ($T \approx 0.7\,\rm MeV$) must
not be too large.  The energy density contribution of a sterile
neutrino is often described as the fraction of the energy density in a
fully populated (thermal) single neutrino-antineutrino species plus
the energy density in the active neutrinos, $N_\nu$. Depending on
one's particular adoption of observationally inferred primordial
abundances, one can arrive at limits between $N_\nu < 3.2$ and $N_\nu
< 4$ \cite{schramm,afs,lisi}.  We have calculated the energy density
contributed by the sterile neutrino at BBN, and its limits are shown
in Figs.\ \ref{L0} and \ref{Ln0}.

\subsection{$^6$Li and D photoproduction}
Another constraint arises from photoproduction of deuterium (D) and
$^6$Li stemming from the decay of massive sterile neutrinos after BBN
\cite{photoprod}.  Energetic cascades dissociate $^4$He into excessive
amounts of D, which is bounded observationally \cite{burtyt}.  Also,
energetic $^3$H and $^3$He produced in the cascades can synthesize
$^6$Li through $^3{\rm H}(^3{\rm He}) + ^4{\rm He} \rightarrow ^6{\rm
Li} + {\rm n(p)}$, overproducing $^6$Li, which is also bounded
observationally \cite{li6obs}.

\begin{figure}[ht]
\includegraphics[width=3.3truein]{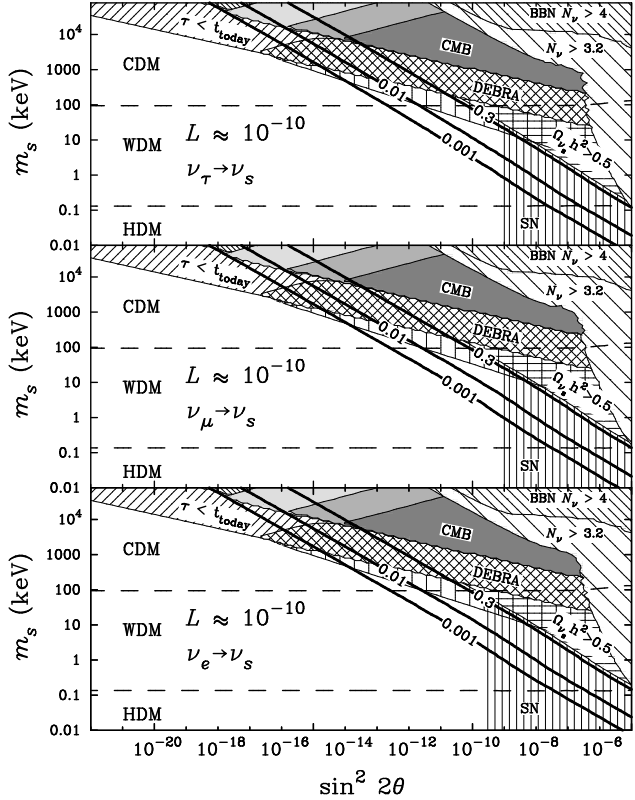}
\caption
{\small Contours of closure fraction are displayed as thick lines
labelled with $\Omega_{\nu_s} h^2$ for negligible lepton asymmetry
($L\approx 10^{-10}$) for the three neutrino flavors mixing with a
massive sterile neutrino.  Shown are constraints from the energy
density present at CMB decoupling, with the dark gray region
corresponding to BOOMERanG/MAXIMA's present limits.  The medium gray
region corresponds to MAP's predicted future constraint, and the light
gray region is the Planck mission's potential constraint region.  The
constraints from DEBRA, and BBN are so labelled. The region of
potential constraints from supernovae (SN) is also shown as closely
spaced vertical lines.  The region of widely spaced vertical lines is
where radiative decays today may give detectable X-ray
signatures. Regions filled with horizontal lines are where
$\Omega_{\nu_s} h^2 > 0.5$ and are inconsistent with the observed age
of the universe. Sterile neutrinos with parameters in the region
labelled by ``$\tau < t_{\rm today}$'' decay without constrainable
effects and make no contribution to the present matter density. }
\label{L0}
\end{figure}

\begin{figure}[ht]
\includegraphics[width=3.3truein]{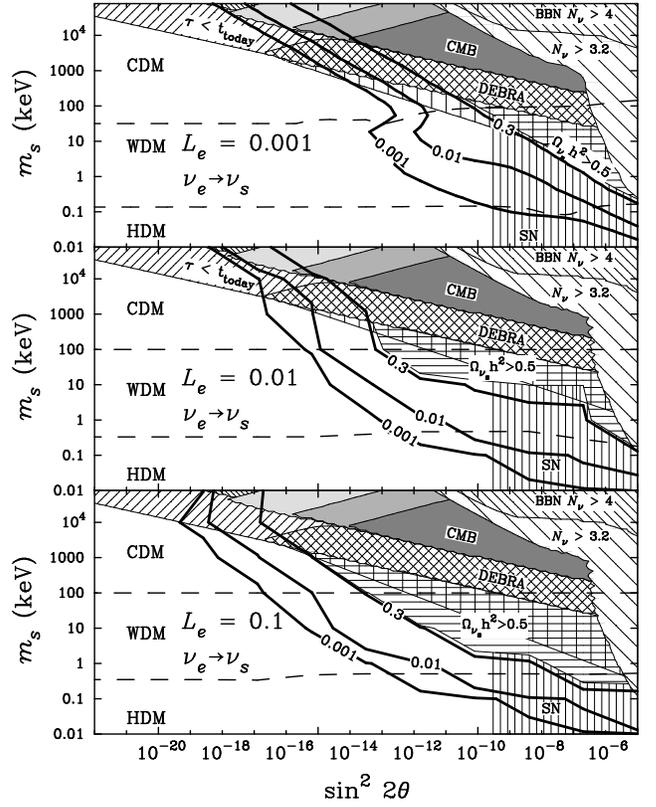}
\caption
{\small Contours of closure fraction are displayed as thick lines
labelled with $\Omega_{\nu_s} h^2$ for $L=0.001,0.01,0.1$. Constraints
are shown as in Fig.\ \ref{L0}.}
\label{Ln0}
\end{figure}

These constraints lie in regions of parameter space which are almost
identical to those bounded by CMB considerations.  The $^6$Li limit
can be seen to extend to the small region just to the left of the CMB
constraints in Figs. \ref{L0} and \ref{Ln0} (closely hatched area,
hatching oriented from the lower right to the upper left).  It should
be noted that the $^6$Li/D photoproduction constraints are from
current observation, while the potential constraints from MAP and
Planck of the energy density present at CMB decoupling are a {\it
future} possibility that can corroborate the photoproduction
constraints.

\section{Active-Sterile Neutrino Transformation in Core-Collapse Supernovae}
\label{supernovaconstraints}

In the previous sections, we have calculated the ranges of sterile
neutrino masses and mixing angles (and primordial lepton asymmetries)
for which sterile neutrinos can account for some or all of the dark
matter.  In this section we describe some of the implications for
core-collapse (Type Ib/c, II) supernovae of active-sterile neutrino
transformation with parameters in these regions.  Neutrinos play a
dominant, pervasive role in core-collapse supernovae, so the stakes
are high whenever we introduce non-standard neutrino physics like
neutrino mixing. 

In broad brush the effects are simple: too much neutrino conversion in
a supernova results in too much energy loss to sterile neutrinos,
manifestly in conflict with observation (the mixing angles considered
here are sufficiently small that the sterile neutrinos are not trapped
in the core). Requiring the hemorrhaging in sterile neutrinos not to
be too great then places bounds on neutrino mixing parameters.

In fine detail, however, the procedure is not so simple, since the
coupled supernova neutrino transport-transformation problem is highly
nonlinear and the dynamics very difficult to treat analytically or
numerically. Confounding the matter is our relatively poor
understanding of the physics of the core-collapse phenomenon itself.

Nevertheless, we have boldly attempted (as have several workers before
us \cite{super,super2,dhrs2,dh,mutaucore}) to describe some of the
salient effects of neutrino transformation in a supernova and have
extracted some rather conservative limits on mixing parameters,
conservative, at least, for arguing the case for sterile neutrino dark
matter.  Our ``limits'' are by no means final, and much future work
remains to be done in order ascertain the viability of sterile
neutrino dark matter.  In what follows, we give a brief biography of a
core-collapse supernova, describe the relevant neutrino transport and
transformation physics, and indicate how we obtained the supernova
``constraint'' regions in Figs.~\ref{omegaepanel},
\ref{omegataupanel}, \ref{mfsfig}, \ref{L0}, and \ref{Ln0}.

Core-collapse supernovae are the death throes of massive stars. A star
of mass $\gtrsim 10 M_\odot$ dies when its $\approx 1.4 M_\odot$ iron
core undergoes gravitational collapse to a neutron star, an event
releasing $\approx 99\%$ of the $\approx 10^{53} {\rm\,erg}$
gravitational binding energy of the neutron star in all active species
of neutrinos (see Refs.~\cite{shapiroteuk,bethermp} for an
introduction). In the standard picture this event lasts $\approx 10-15
{\rm\,s}$ and can be divided roughly into three phases: the infall,
shock reheating, and $r$-process epochs.

During the infall epoch, the iron core collapses on a near
gravitational time scale of $\sim 1{\rm\,s}$ to a dense
proto-neutron star. As the core density rises, the forward reaction in
\begin{equation}
e^- + \text{``}p\text{''} \rightleftharpoons \nu_e + \text
{``}n\text{''}
\label{neutronization}
\end{equation}
neutronizes the core until the neutrino trapping density $\rho \sim
10^{11}-10^{12} {\rm\,g\,cm^{-3}}$ is reached, a condition which
allows the forward and reverse reactions in Eq.\ (\ref{neutronization})
to achieve $\beta$-equilibrium. This equilibrium is established with
relativistically degenerate electrons ($\mu_e \approx 25 - 220
{\rm\,MeV}$) and electron neutrinos ($\mu_{\nu_e} \approx 10 - 170
{\rm\,MeV}$).

The quotes in Eq.\ (\ref{neutronization}) refer to free {\it and} bound
baryons. Most of the baryons are bound in large nuclei, since the
infall-collapse phase is characterized by temperatures $T\approx
1-3{\rm\,MeV}$ and a low entropy-per-baryon $s\approx 1.5$ (in units
where Boltzmann's constant is unity).

The infall epoch ends when the core density reaches and exceeds the
saturation density of nuclear matter. The inner core bounces, yielding
an outward-propagating shock wave at its boundary with the outer
core. The shock loses much of its energy dissociating the nuclei in
the outer core and mantle of the pre-supernova star and eventually
stalls some $500 {\rm\,km}$ from the center of the core, far short of
generating a supernova explosion. The post-bounce core is a hot, dense
proto-neutron star consisting of free baryons, electrons and positrons,
and active neutrino-antineutrino pairs (and perhaps also muon-antimuon
pairs and strange quark matter), all in thermal and chemical
equilibrium at a temperature $T \approx 30-70 {\rm\,MeV}$ and
entropy per baryon $s\approx 10$.

During the shock reheating epoch, at times post core bounce
$t_{\rm pb}\approx 0.06-1 {\rm\,s}$, neutrinos diffuse out of the
core and deposit energy behind the stalled shock, driving the supernova
explosion \cite{bethewilson}.

In the $r$-process epoch neutrinos continue to diffuse out of the star
at times $t_{\rm pb}\approx 1-15 {\rm\,s}$. The neutrinos drive mass
loss from the proto-neutron star, possibly setting the physical
conditions for r-process (heavy element) nucleosynthesis
\cite{rprocess,neutwind}.

The star's nuclear composition becomes increasingly neutron rich as
the electron neutrinos depart the star and the reactions in
Eq.\ (\ref{neutronization}) proceed with continuous, local
re-establishment of $\beta$ equilibrium. For example, at core bounce
($t_{\rm pb} = 0$), the electron fraction (net number of electrons per
baryon) in the center of the core is $Y_e\approx 0.35$; a cold neutron
star, the end point of evolution in this case, has $Y_e \lesssim 0.01$.

Since neutrinos dominate the energetics of core-collapse (Type Ib/c,
II) supernovae, appreciable active-sterile neutrino transformation can
completely alter the standard picture of stellar collapse. Although
this picture has been refined by observations of neutrinos from
SN 1987A \cite{sn1987a}, large-scale numerical simulations
\cite{bethewilson, snnumerical}, and semi-analytical work
\cite{bethewilson, betheemp, bbal, fullerthesis}, it is not clear that
this is the only way for supernovae to evolve. Nevertheless, by
requiring that not too much energy be ``lost'' too quickly to singlet
neutrinos in the proto-neutron star, and hence avoiding a conflict
with standard supernova theory and observations, we can delimit
regions in the $m_s-\sin^2 2\theta$ plane which may adversely affect
its evolution.  In fact, many supernova constraints on new physics
rely on limiting the energy loss to exotic particles in a
proto-neutron star \cite{raffelt}.

Sterile neutrinos can be produced in supernovae either coherently
through mass level-crossings or incoherently via scattering-induced
wave function collapse. Whether the former or latter process dominates
depends on the hierarchy of length scales relevant for neutrino
transport and transformation: the local neutrino mean free path
$\lambda$, local neutrino oscillation length in matter $l_m$, and
resonance width $\delta r$ [the supernova analogue of Eq.\
(\ref{reswidth})]. Neutrino flavor eigenstate evolution is coherent
and transformation is uninterrupted through resonance if $\lambda \gg
\delta r$. Neutrino flavor transformation is also adiabatic if $\delta
r \gg l_m^{\rm res}$, where $l_m^{\rm res}\ge l_m$ is the oscillation
length at resonance. This is in complete analogy to the evolution
through resonances discussed above in Sec. \ref{leptressect} for the
early universe. If instead neutrino conversion is incoherent and
transformation is interrupted often by collisions, then the conversion
will be suppressed by quantum damping if $l_m > \lambda$.

It is easy to show that incoherent conversion dominates sterile
neutrino production in supernova cores.  In a proto-neutron star of
radius $R=10-50 {\rm\,km}$ the density varies relatively slowly with
distance from the center of the star, so the scale height of the weak
potential is $H = |d\ln V/dr|^{-1} \approx 10-100 {\rm\,km}$. Then the
resonance width is $\delta r = H \tan 2\theta \approx (10^6 - 10^7
{\rm\,cm}) \tan 2\theta$. The mean free path of a typical neutrino is
about $\lambda \approx 10 {\rm\,cm}$ (actually $\lambda \sim
10{\rm\,cm}\text{ to }\sim 10\rm\, m$, but this range makes little
difference for our conclusions). Therefore, neutrino evolution is
coherent if $\sin^22\theta \lesssim 10^{-12} - 10^{-10}$, and the
maximal resonance width giving coherent evolution is $\delta r_{\rm
max} \approx 10 {\rm\,cm}$. Now the oscillation length at resonance is
$l_m^{\rm res}=4\pi E/\delta m^2 \sin 2\theta \approx 10^{-4}
{\rm\,cm\,} (10 {\rm\,keV}/m_s)^2/\sin 2 \theta$, so coherent
evolution gives $l_m^{\rm res} \gtrsim 10-100 {\rm\,cm\,} (10
{\rm\,keV}/m_s)^2$. In these conditions, however, the evolution is not
adiabatic, since $\delta r_{\rm max} \lesssim l_m^{\rm res}$. As a
result, mass level crossings contribute only subdominantly to sterile
neutrino production in supernovae. Conversion to sterile neutrinos
less massive than $10 {\rm\,keV}$ is even less efficient, and more
massive neutrinos will not have resonances.

Incoherent neutrino production is weakly damped at (or away from) a
resonance when $\sin^22\theta \gtrsim 10^{-10} (10 {\rm\,keV}/m_s)^4$.
The local oscillation length in matter away from resonance is
typically much smaller than the oscillation length at resonance, so
for sterile neutrino masses $m_s \gtrsim 10 {\rm\,eV}$ weak damping
generally obtains off resonance even if the condition above is not
met. Furthermore, since resonances last only a fraction $\lesssim
10^{-8}$ of the neutrino diffusion time scale, as derived below, most
neutrino conversion in a core-collapse supernova occurs in the weak
damping regime.

The time rate of energy ``loss'' $\cal E$ to sterile neutrinos (and
sterile antineutrinos) per unit mass in a proto-neutron star is
proportional to the active neutrino energy, scattering cross section
on weak targets, and the average conversion probability
\cite{super,super2,raffelt}:
\begin{align}
{\cal E} &\approx \frac{1}{m_N}\int{d\Phi_{\nu_\alpha} E\,
\sigma_{\nu_\alpha}(E) \frac{1}{2} \langle
P_m(\nu_\alpha\rightarrow\nu_s; p,t) \rangle}\nonumber\\
&\quad + 
\frac{1}{m_N}\int{d\Phi_{\bar{\nu}_\alpha} E\,
\sigma_{\bar{\nu}_\alpha}(E) \frac{1}{2} \langle
P_m(\bar{\nu}_\alpha\rightarrow\bar{\nu}_s; p,t) \rangle},
\label{emissivity}
\end{align}
where, for the sake of simplicity, we have suppressed Pauli-blocking
effects. In Eq.\ (\ref{emissivity}), the cross sections for neutrino
scattering on free baryons are
$\sigma_{\nu_\alpha}(E)\approx\sigma_{\bar{\nu}_\alpha}(E)\approx
1.66G_F^2 E^2$, and the differential neutrino and antineutrino fluxes
are
\begin{eqnarray}
d\Phi_{\nu_\alpha} &=& c\ dn_{\nu_\alpha} \approx \frac{d^3 p}{(2\pi)^3}
\frac{1}{e^{E/T_{\nu_\alpha} - \eta_{\nu_\alpha}}+1}
\nonumber\\
&\approx&
\frac{1}{2\pi^2}\frac{E^2 dE}{e^{E/T_{\nu_\alpha} - \eta_{\nu_\alpha}}+1}
\label{supernovafluxnu}
\end{eqnarray}
\begin{eqnarray}
d\Phi_{\bar{\nu}_\alpha} &=& c\ dn_{\nu_\alpha} \approx \frac{d^3 p}{(2\pi)^3}
\frac{1}{e^{E/T_{\bar{\nu}_\alpha} -
\eta_{\bar{\nu}_\alpha}}+1}
\nonumber\\
&\approx&
\frac{1}{2\pi^2}\frac{E^2 dE}{e^{E/T_{\bar{\nu}_\alpha} -
\eta_{\bar{\nu}_\alpha}}+1}
\label{supernovafluxnubar}
\end{eqnarray}
for relativistic neutrinos, where $\eta_{\nu_\alpha} =
\mu_{\nu_\alpha}/T_{\nu_\alpha}$ is the denegeracy parameter of
$\nu_\alpha$ as above. A simple but somewhat crude criterion for
avoiding conflict with supernova theory and observations of SN 1987A is
${\cal E} \lesssim 10^{19} {\rm\,erg\,s^{-1}\,g^{-1}}$ \cite{raffelt},
a limit on the sterile neutrino emissivity equivalent to a loss of
$\sim 10 {\rm\,MeV}$ per baryon per second.

The weak potentials [cf. Eqs.\ (\ref{generalvd}) and (\ref{vt})] which drive
neutrino flavor transformation take a slightly different form in
supernovae than in the early universe \cite{nr}. For
$\nu_e\rightleftharpoons\nu_s$ transformation the potential stemming
from finite density effects is [from Eq.\ (\ref{generalvd})]
\begin{equation}
V^D = \frac{G_F\rho}{\sqrt{2}m_N}\left(3Y_e - 1 + 4Y_{\nu_e} +
2Y_{\nu_\mu} + 2Y_{\nu_\tau}\right),
\label{supernovaVLe}
\end{equation}
where $m_N\approx 931.5 {\rm\,MeV}$ is an atomic mass unit, $n_i$ is
the number density of species $i$, and $Y_i \equiv (n_i-n_{\bar{i}}) /
n_B$ is the net fraction of species $i$ per baryon ($n_B =
n_n+n_p$). In part, the form of Eq.\ (\ref{supernovaVLe}) follows from
local charge neutrality and the assumption that muons, anti-muons, and
strange quark matter are negligibly populated. The finite density
potential for $\nu_\mu\rightleftharpoons\nu_s$ transformation is [from
Eq.\ (\ref{generalvd})]
\begin{equation}
V^D = \frac{G_F\rho}{\sqrt{2}m_N}\left(Y_e - 1 + 2Y_{\nu_e} +
4Y_{\nu_\mu} + 2Y_{\nu_\tau}\right).
\label{supernovaVLmu}
\end{equation}
The potential for $\nu_\tau\rightleftharpoons\nu_s$ transformation
follows from Eq.\ (\ref{supernovaVLmu}) upon switching the labels
$\nu_\mu$ and $\nu_\tau$. The finite temperature potential $V^T$ is
negligible in supernovae \cite{nr}, so the total potentials
for neutrino and antineutrino transformation are $V = V^D + V^T
\approx V^D$ and $\bar{V} = -V^D + V^T \approx -V^D \approx -V$,
respectively.

The average oscillation probabilities in Eq.\ (\ref{emissivity})
depend on the weak potentials and neutrino mixing parameters, from
Eqs.\ (\ref{avgprob}), (\ref{avgprobnubar}):
\begin{align}
&\langle P_m(\nu_\alpha\rightarrow\nu_s; p.t) \rangle \approx \nonumber\\
&\quad \frac{1}{2}
\frac{\Delta(E)^2 \sin^2
2\theta}{\Delta(E)^2\sin^2 2\theta + D^2 +
[\Delta(E) \cos 2\theta - V^D]^2}
\label{supernovaavgprob}
\\
&\langle P_m(\bar{\nu}_\alpha\rightarrow\bar{\nu}_s; p,t) \rangle 
\approx \nonumber\\
&\quad\frac{1}{2} \frac{\Delta(E)^2 \sin^2 2\theta}{\Delta(E)^2\sin^2
2\theta + \bar{D}^2 + [\Delta(E) \cos 2\theta + V^D]^2},
\label{supernovaavgprobnubar}
\end{align}
where $\Delta(p) = \delta m^2/2p \approx m_s^2/2E \approx \Delta(E)$
for relativistic neutrinos. As before the quantum damping rate $D =
\Gamma_{\nu_\alpha}/2 = \int{d\Phi_{\nu_\alpha}
\sigma_{\nu_\alpha}(E)}/2$ for neutrinos is one-half the neutrino
scattering rate; an analogous expression applies for the antineutrino
damping rate $\bar{D}$.  The coherent effects described in Ref.\
\cite{dolgovnonres} may modify these conversion rates.

The conditions for resonance $\pm V^D=\Delta(p) \cos 2\theta \approx
m_s^2 \cos 2\theta/2E$, the potentials in
Eqs.\ (\ref{supernovaVLe}), (\ref{supernovaVLmu}), and the average
oscillation probabilities in
Eqs.\ (\ref{supernovaavgprob}), (\ref{supernovaavgprobnubar}) imply a
negative feedback between active (anti)neutrino transformation and the
potentials. If quantum damping effects are comparable for neutrinos
and antineutrinos or are relatively unimportant, as argued above for
most of the parameter range of interest, neutrino conversion is
enhanced relative to antineutrino conversion when $V^D>0$. The
preferential conversion of neutrinos {\it decreases} the potential,
since the finite density part of the potential depends on the relative
excess (or deficit) of particles over antiparticles (the net
contribution of particle-antiparticle pairs is zero). As the potential
decreases, the neutrino and antineutrino conversion rates approach the
common rate they would have at vanishing potential. Since neutrinos
and antineutrinos of all flavors are produced in roughly equal numbers
and their individual number densities are at least comparable to the
electron density, the dynamical feedback mechanism may drive the
potential $V^D$ locally to zero throughout the proto-neutron star, as
long as the time scale for this process is less than the local
neutrino diffusion time scale $t_d \sim \lambda (R/\lambda)^2/c \sim 3
{\rm\,s}$; a similar sequence will ensue if $V^D<0$.  If this happens,
the equal neutrino and antineutrino conversion rates ensure that the
potential will remain zero, with any local deviation continuously
smoothed out, up to the higher order effects of neutrino diffusion. 

In this sense zero potential is a fixed point of the dynamics of
neutrino transport and conversion in core-collapse supernovae and the
time scale for achieving zero potential is a dynamical equilibration
time scale. The situation is more complicated if sterile neutrinos mix
with multiple active neutrinos or if damping is important. For
example, an excess of neutrinos over antineutrinos implies damping is
asymmetric, and this can retard equilibration. These issues will be
examined in more detail in a separate work \cite{sterilecore}.  We
note in passing that this behavior can also occur when neutrino
scattering is rare and mass level crossings dominate sterile neutrino
production \cite{activesterilerproc}; some of this physics has also
been discussed in connection with active-active neutrino
transformation in proto-neutron stars \cite{mutaucore}.

We can estimate the equilibration time scale $\tau_V$ for driving a
potential $V^D_0$ to zero by computing the time over which the
disparity in the rates of neutrino and antineutrino conversion will
induce a potential $\delta V^D = -V^D_0$ which cancels the original
potential $V^D_0$. Now Eqs.\ (\ref{supernovaVLe}), (\ref{supernovaVLmu})
imply $\delta V^D = 2\sqrt{2}G_F(\delta n_{\nu_\alpha}-\delta
n_{\bar{\nu}_\alpha})$, where $\delta n_{\nu_\alpha}$ and $\delta
n_{\bar{\nu}_\alpha}$ are the numbers of neutrinos and antineutrinos
converted per unit volume, respectively. The number densities of
(anti)neutrinos converted are proportional to the rates of
(anti)neutrino conversion per baryon, the baryon density, and the time
over which the conversion takes place.  If we assume $\tau_V$ is small
compared to the time scales for the change of the density and
temperature in the star (a ``static'' approximation), we have $\delta
n_{\nu_\alpha} = -n_B \tau_V \int{d\Phi_{\nu_\alpha}
\sigma_{\nu_\mu}(E) \frac{1}{2} \langle
P_m(\nu_\mu\rightarrow\nu_s; p,t)\rangle}$ and similarly for $\delta
n_{\bar{\nu}_\alpha}$. Solving for $\tau_V$ and using $n_B = \rho/m_N$
gives the time scale for achieving dynamic equilibrium in the
$\nu_\alpha\rightleftharpoons\nu_s$,
$\bar{\nu}_\alpha\rightleftharpoons\bar{\nu}_s$ system:
\begin{align}
\tau_V &= \frac{V^D_0 m_N}{2\sqrt{2}G_F\rho}\Bigg(\int{d\Phi_{\nu_\alpha}
\sigma_{\nu_\alpha}(E) \frac{1}{2} \langle
P_m(\nu_\alpha\rightarrow\nu_s; p,t)\rangle} \nonumber\\
&\quad-\int{d\Phi_{\bar{\nu}_\alpha}
\sigma_{\bar{\nu}_\alpha}(E) \frac{1}{2} \langle
P_m(\bar{\nu}_\alpha\rightarrow\bar{\nu}_s; p,t)\rangle}\Bigg)^{-1}.
\label{supernovatauv}
\end{align}

It is instructive to evalute the equilibration time scale $\tau_V$ for
the $\nu_\mu\rightleftharpoons\nu_s$,
$\bar{\nu}_\mu\rightleftharpoons\bar{\nu}_s$ system in two limits: (i)
far away form a resonance, for which $|V^D_0| \gg
\Delta(E)\cos2\theta$; and (ii) very near or at a resonance, where
$|V^D_0|\approx \Delta(E)\cos2\theta\approx m_s^2/2E$ for $\theta \ll
1$. In the off-resonance case, inserting
Eqs.\ (\ref{supernovafluxnu})-(\ref{supernovafluxnubar}) and
Eqs.\ (\ref{supernovaavgprob})-(\ref{supernovaavgprobnubar}) in
Eq.\ (\ref{supernovatauv}) and expanding to lowest order yields
\begin{align}
\tau^{\rm off-res}_V &\approx
\frac{96}{1.66\sqrt{2}}\frac{\left(V^D_0\right)^4 m_N}{G_F^3 \rho T^2
m_s^6 \sin^22\theta} \nonumber\\ &\approx \frac{1.5\times
10^{-8}{\rm\,s}}{\sin^22\theta}
\left(\frac{10^{14}{\rm\,g\,cm^{-3}}}{\rho}\right)\nonumber\\
&\quad\times\left(\frac{50{\rm\,MeV}}{T}\right)^2
\left(\frac{10{\rm\,keV}}{m_s}\right)^6
\left(\frac{V^D_0}{1{\rm\,eV}}\right)^4,
\label{tauvoffresonance}
\end{align}
where the factor of $1.66$ comes from the neutrino scattering
cross section given earlier. We have derived
Eq.\ (\ref{tauvoffresonance}) by assuming a locally constant density
and temperature, consistent with our ``static'' approximation. We have
also ignored the build up of a non-zero muon neutrino chemical
potential $\mu_{\nu_\mu}$ and accompanying Pauli blocking in neutrino
scattering and pair production as equilibration proceeds, a
simplification giving $d\Phi_{\nu_\mu}\approx
d\Phi_{\bar{\nu}_\mu}$. Using the same simplifications, we can find
the time scale in the on-resonance case:
\begin{align}
\tau^{\rm on-res}_V &\approx \frac{8\pi^2}{(1.66)45\sqrt{2}\zeta(5)}
\frac{|V^D_0| m_N}{G_F^3 \rho T^5} \nonumber\\
&\approx
\frac{16\zeta(3)}{(1.66)7\pi^2\sqrt{2}\zeta(5)} \frac{m_s^2 m_N}{G_F^3
\rho T^6} \nonumber \\ 
&\approx 2 \times 10^{-9} {\rm\,s}
\left(\frac{50{\rm\,MeV}}{T}\right)^5
\left(\frac{10^{14}{\rm\,g\,cm^{-3}}}{\rho}\right)
\left\vert\frac{V^D_0}{1{\rm\,eV}}\right\vert\nonumber\\
&\approx 6.6 \times 10^{-10}
{\rm\,s}
\left(\frac{50{\rm\,MeV}}{T}\right)^6
\left(\frac{10^{14}{\rm\,g\,cm^{-3}}}{\rho}\right)\nonumber\\
&\quad\times\left(\frac{m_s}{10 {\rm\,keV}}\right)^2.
\label{tauvonresonance}
\end{align}
The second and fourth expressions of Eq.\ (\ref{tauvonresonance})
follow from the resonance condition $|V^D_0| \approx m_s^2/2E$ and
taking the neutrino energy $E$ to be the average neutrino energy
$\langle E\rangle = \int{d\Phi_{\nu_\mu} E} / \int{d\Phi_{\nu_\mu}}
\approx 7\pi^4T_{\nu_\mu}/180\zeta(3)\approx 3.15T_{\nu_\mu}$ in the
proto-neutron star.

The on- and off-resonance equilibration time scales both vary
inversely with the ambient density and temperature. Hotter conditions
enhance the neutrino conversion rate per scatterer, and denser
conditions enhance the number density of scatterers. In either case it
takes less time to drive a pre-existing potential to zero. On the
other hand, only the off-resonance time scale $\tau^{\rm off-res}_V$
depends on the vacuum mixing angle; the inverse dependence is natural,
since the emissivity in sterile (anti)neutrinos is proportional to
$\sin^22\theta$. The on-resonance time scale $\tau^{\rm on-res}_V$ is
independent of $\sin^22\theta$, because the relevant mixing angle at a
resonance is $\pi/4$. If $V^D_0>0$ neutrino conversion dominates
antineutrino conversion at resonance and vice versa if $V^D_0<0$, so
this time scale is extremely short compared to the neutrino diffusion
time scale or indeed {\it any} time scale typically associated with a
core-collapse supernova. Individual resonances are fleeting and result
in negligible energy loss to sterile neutrinos. As a result, we may
safely assume that most neutrino transformation takes place
off resonance, with an equilibration time scale given by
Eq.\ (\ref{tauvoffresonance}).

We can delimit ranges of the neutrino mixing parameters $m_s$
and $\sin^22\theta$ which {\it may} adversely affect core-collapse
supernovae by evaluating the sterile neutrino emissivity in
Eq.\ (\ref{emissivity}) in the following manner.

In the case of $\nu_e\rightleftharpoons\nu_s$ transformation, the
potential $V^D$ is initially positive in the post-bounce supernova
core. If the neutrino mixing parameters happen to give $\tau^{\rm
off-res}_V > t_d$, core neutronization proceeds roughly on a diffusion
time scale, although the conversion of electron neutrinos accelerates
the shift of $\beta$ equilibrium toward
neutron richness. Neutronization decreases the electron fraction
$Y_e$, and neutrino conversion decreases the electron neutrino
fraction $Y_{\nu_e}$. As a consequence, the potential decreases and
eventually reaches a value such that $\tau^{\rm off-res}_V < t_d$.

Once this occurs, neutrino conversion quickly resets the potential to
zero and maintains this value as neutronization continues. If instead
the neutrino mixing parameters happen to give $\tau^{\rm off-res}_V <
t_d$ at core bounce, conversion immediately resets and maintains the
potential at zero.

In the case of $\nu_\mu\rightleftharpoons\nu_s$ or
$\nu_\tau\rightleftharpoons\nu_s$ transformation, the potential $V^D$
is initially {\it negative} in the post-bounce core. Here conversion
has no direct effect on the rate of neutronization (notwithstanding
feedback on the nuclear equation of state), which increases the
magnitude of the potential as the electron fraction falls to a value
$\sim 0.01$. Unless $\tau^{\rm off-res}_V < t_d$ from the outset,
neutronization dictates the evolution of the potential.

The emissivity ${\cal E}$ depends on the local, instantaneous value of
the potential, so the total energy lost to sterile neutrinos depends
on the time history of the spatially varying potential and, in
particular, whether neutronization or neutrino conversion locally
dominates its evolution. An accurate estimate of the
spatially integrated emissivity and, hence, of the regions in
parameter space which may alter the standard picture of stellar
collapse clearly requires a detailed investigation of the dynamics of
these systems, as well as a better understanding of the physics of
core-collapse supernovae.

From the viewpoint of the viability of sterile neutrinos as dark
matter, however, we may {\it conservatively} estimate the emissivity
by assuming the system achieves zero potential relatively quickly, so
that ${\cal E}$ is evaluated with $V^D=0$. Evaluating Eq.\
(\ref{emissivity}) in this limit and imposing the condition ${\cal E}
\lesssim 10^{19} {\rm\,erg\,s^{-1}\,g^{-1}}$ gives the disfavored
regions $\sin^22\theta \gtrsim 3\times 10^{-10}$ for
$\nu_e\rightleftharpoons\nu_s$ transformation and $\sin^22\theta
\gtrsim 10^{-9}$ for $\nu_\mu\rightleftharpoons\nu_s$ or
$\nu_\tau\rightleftharpoons\nu_s$ transformation for sterile neutrino
masses $m_s\gtrsim 10 {\rm\,eV}$.  For smaller masses, the limits are
significantly weaker, owing to the onset of quantum damping.  These
limits are shown in Figs.~\ref{omegaepanel}, \ref{omegataupanel},
\ref{mfsfig}, \ref{L0}, and \ref{Ln0}. The emissivity falls with
increasing potential away from a resonance, so more detailed future
work on this complex problem may find actual constraints which are
weaker than the very conservative limits given here.

Sterile neutrino dark matter parameters which lie near the edges of
the disfavored regions in Figs.~\ref{omegaepanel},
\ref{omegataupanel}, \ref{mfsfig}, \ref{L0}, and \ref{Ln0} could give
interesting signals in current and future supernova neutrino detectors
(see, {\it e.g.}, Ref.~\cite{beacomsuper}). These detectors possibly
could discern unique signatures for sterile neutrinos. Such signatures
would bolster the case for sterile neutrino dark matter.

To summarize this section, we have delimited conservatively the
regions in the parameter space of active-sterile neutrino mixing that
are disfavored by energy-loss considerations in core-collapse
supernovae. We have found that the coupled problem of neutrino
transport and flavor transformation in hot and dense nuclear matter is
a formidable one. It involves following the local evolution of the
weak potential which drives flavor transformation, including the
feedback from diffusion and the conversion itself. We have described
this physics roughly by estimating the competing time scales for
lepton number diffusion and for cancellation of the
potential. Depending on the mixing parameters and the spacetime
evolution of the potential, the potential may well be reset to zero in
much of the proto-neutron star, so the effects of neutrino propagation
in matter need not suppress sterile neutrino (or antineutrino)
production. As a result, core-collapse supernovae can be significantly
more sensitive to active-sterile neutrino mixing than they were found
to be in previous studies \cite{super,dhrs2,dh}. These studies used a
spatially and temporally constant value of the potential, resulting in
limits on mixing angles similar to ours but for significantly larger
sterile neutrino masses $\gtrsim 10{\rm\,keV}$; for smaller masses
their limits on mixing angles are weaker because the putative matter
effects suppress conversion \cite{super,dhrs2,dh}. 

Of course, we have deliberately chosen to be conservative in applying
supernova limits, since our objective in this work is to assess the
viability of sterile neutrino dark matter. The true limits, quite
interesting in their own right and obtained from a self-consistent and
proper treatment of the full, multi-dimensional Boltzmann evolution of
the neutrino seas coupled with the nuclear equation of state, may well
lie somewhere in between the values determined in this and previous
studies. We leave attempts at such investigations for a future work
\cite{sterilecore}.

\section{Conclusions}
\label{conclusions}

We have estimated the resonant and non-resonant scattering prodution
of sterile neutrinos in the early universe. The basis for the
production of these sterile species is a presumed mixing with active
neutrinos in vacuum. Of course, such mixing renders these species not
truly ``sterile.'' As a result, the ``sterile'' neutrinos can decay
and this, together with their overall contribution to energy density,
constitutes the basis for several stringent cosmological and astrophysical
constraints which we have discussed in detail.

Additionally, these sterile species may produce significant effects in
core-collapse supernovae. Some of these effects, such as massive core
energy loss, could be the basis for true constraints. However, it must
be kept in mind that (1) the supernova explosion energy is only some
$\sim 1\%$ of the total energy resident in the active neutrino seas
and that (2) we do not yet understand in detail how supernovae explode
(nor do we have a sufficiently detailed observed core-collapse
neutrino signal to place stringent constraints). Better theoretical
understanding of supernova physics, perhaps coupled with the neutrino
signature of a Galactic core-collapse event, may allow the indicated
regions on Figs.~\ref{L0} and \ref{Ln0} to become true hard and fast
constraints instead of simply ``disfavored'' parameter regions. In
fact, deeper insight into the time evolution of the potentials
governing sterile neutrino production in the core may allow {\it
extension} of the constrained parameter region to even smaller values
of vacuum mixing angle.

Nevertheless, it is clear from our work that sterile ``neutrino''
species with ranges of masses and vacuum couplings could be produced
in quantities sufficient to explain {\it all} of the non-baryonic dark
matter, while evading all present day laboratory and astrophysical
constraints. Within the allowed ranges in mass and mixing parameters
which give these dark matter solutions are regions where the sterile
neutrino masses and/or energy spectra combine to produce collisionless
damping scales corresponding to warm or even cold dark matter,
subsuming the interesting behavior range for large scale structure.

It is then a disturbing possibility that the dark matter might not be
``weakly interacting massive particles'' (WIMPs), but rather ``nearly
non-interacting massive particles'' (NNIMPs) which are likely not
detectable in ordinary dark matter detection experiments. This
possibility begs two questions: (1) how could we hope to constrain or
definitively detect this dark matter candidate; and (2) what are
sterile neutrinos?

The answer to the first question is more straightforward than the
resolution of the second. As outlined above, better understanding of
the neutrino and equation of state physics of core-collapse supernovae
could help us extend constraints. Improved observations and models of
x-ray emission from clusters of galaxies and other objects could
provide stringent constraints or, conceivably, even detections of
sterile neutrino photon decay.

As discussed above, future observations bearing on the clustering of
dwarf galaxy or Lyman-$\alpha$ cloud halos at high redshift may
provide evidence for or against WDM as opposed to CDM. Direct evidence
for WDM would constitute a point in favor of sterile neutrinos, though
not a definitive one by any means. Likewise, and insidiously, our
allowed parameter space for dark matter accommodates standard CDM
behavior, even for lepton numbers equal to the baryon number.

Direct detection of WIMPs in the laboratory or detection of gamma rays
associated with WIMP annihilation in galactic centers
\cite{silksalati} obviously rule out sterile neutrino dark matter,
given the small vacuum mixing angles suggested by our work. It is
worth considering whether $\beta$-decay electron energy spectrum or
pion decay experiments could be pushed in sensitivity to the point
where massive sterile neutrinos in some of the allowed regions of
Figs.~\ref{L0} and \ref{Ln0} could be constrained. This would require
an increase in sensitivity to $m_s\sin^22\theta$ of at least some six
orders of magnitude and this is clearly untenable with {\it current}
technology \cite{behrpersonal}.  Finally, although a
number of extensions of the standard model motivate the existence of
multiple sterile neutrinos, it must be pointed out there is no
independent physics suggestion for the sterile neutrino mass ($\sim
1{\rm\,keV}$ to $\sim 10 {\rm\,MeV}$) and mixing
($10^{-17}\le\sin^22\theta\le3\times 10^{-10}$) parameters which give
viable dark matter candidates in our calculations.

As discussed above, sterile neutrino degrees of freedom with
ultra-large masses ({\it e.g.}, of order the standard model
unification scale, or even the top quark mass) are in some sense
``natural,'' at least in the context of a see-saw explanation for the
low masses of active neutrinos. 

There is now a {\it reasonable} chance that new neutrino experiments
scheduled to come to fruition in the next few years [{\it e.g.}, the
Sudbury Neutrino Observatory (SNO) \cite{sno}, KamLAND \cite{kamland},
ORLaND \cite{orland}, K2K \cite{k2k}, MINOS \cite{minos}, ICARUS
\cite{icarus}, and Mini-BooNE \cite{miniboone}] will allow us to
deconvolve the neutrino mass and mixing spectrum.  This would be an
achievement heavy with implications for many aspects of physics and
astrophysics. Will the requirement for sterile neutrinos remain?

It is now true that the interpretation of the current solar,
atmospheric, and accelerator (LSND) data in terms of neutrino flavor
mixing physics demands the introduction of a sterile neutrino with a
rest mass comparable to that of some of the active neutrinos, {\it
i.e.}, light. Necessarily, then, this sterile neutrino is not the dark
matter candidate we speculate on in this work. However, the
unambiguous establishment of the existence of a light sterile neutrino
would expose our ignorance of physics in the neutrino sector in a
stark and dramatic way. On this score, the Mini-BooNE experiment
\cite{miniboone} and the SNO neutral-current experiment \cite{sno} are
the most crucial ones for sterile neutrino dark matter. A confirmation
of the LSND result invites speculation on the existence of more
massive sterile neutrino states.

\acknowledgments

We thank A.B.~Balantekin, J.~Behr, S.~Burles, D.O.~Caldwell,
J.~Frieman, E.~Gawiser, K.~Griest, W.~Haxton, C.~Hogan, E.W.~Kolb,
R.N.~Mohapatra, S.~Reddy, R.~Rothschild, B.~Sadoulet, D.~Tytler, and
J.~Wadsley for useful discussions. M.P. and K.A. would like to
acknowledge partial support from NASA GSRP. This work was
supported in part by NSF Grant PHY-9800980 at UCSD. G.M.F. would like to
thank the Institue for Nuclear Theory, the Physics Department, and the
Astronomy Department of the University of Washington for hospitality.

\appendix
\section*{Appendix: The General Time-Temperature Relation}

In this appendix, we review the calculation of the temperature
evolution in the early universe through periods of varying statistical
weight in relativistic particles, $g$, Eq.\ (\ref{gstar}). The time
derivative of the temperature can be written as \cite{wfh}
\begin{equation}
\frac{dT}{dt} = \frac{dr}{dt}\Bigg/ \frac{dr}{dT}
\label{timetemp}
\end{equation} 
where $r\equiv \ln(R^3)$ and $R$ is the scale factor. The expansion
rate is determined by the Friedmann equation
\begin{equation}
H = \frac{dR}{dt} \frac{1}{R} = \frac{1}{m_{\rm pl}} \sqrt{\frac{8\pi}{3} \rho_{\rm
tot}}
\approx {0.207\,\rm s^{-1}}\ g^{1/2}\ T^2,
\label{hubble}
\end{equation}
where $\rho_{\rm tot} = (\pi^2/30) g T^4$ is the total energy density,
and $T$ is in MeV in the last approximation of Eq.\ (\ref{hubble}).
Therefore, the first half of the time-temperature relation, Eq.\
(\ref{timetemp}), is straightforward: $dr/dt = 3H$. One can get the
second half through the conservation of comoving energy,
\begin{equation}
\frac{d}{dt}\left(\rho R^3\right) + p \frac{d}{dt}\left(R^3\right) = 0
\end{equation}
This can be rewritten into the desired form
\begin{equation}
\frac{dr}{dT} = \frac{d\rho_{\rm tot}}{dT} \left(\rho_{\rm tot} +
p_{\rm tot}\right)^{-1}.
\label{drdtemp}
\end{equation}
With Eqs.\ (\ref{hubble}) and (\ref{drdtemp}) one has the
general temperature evolution.  

In our case the sterile neutrinos can contribute significantly to the
energy density and pressure.  Approximating all species other than the
sterile neutrino to be relativistic, we have $p_\ast \approx 1/3
\rho_\ast$, where $\rho_\ast$ and $p_\ast$ are the energy and pressure
in all particles other that the sterile neutrinos/antineutrinos.
Therefore,
\begin{equation}
\frac{dT}{dr} = \left(\frac{4}{3}\rho_\ast +\rho_s +p_s\right)
\left(\frac{d\rho_\ast}{dT}+\frac{d\rho_s}{dT}\right)^{-1}.
\end{equation}
The rate of change of the standard energy density is
straightforward:
\begin{equation}
\frac{d\rho_\ast}{dT} = \frac{4\pi^2}{30} g T^3 + \frac{\pi^2}{30}T^4
\frac{dg}{dT}.
\end{equation}
The temperature derivative of the sterile neutrino energy density is 
\begin{eqnarray}
\frac{d\rho_s}{dT} &=& \frac{2}{\pi^2}\int{\left(f_{\nu_s}(p)+
f_{\bar\nu_s}(p)\right) \left[(p^2+m_s^2)/T^2\right]^{1/2}
p^2dp}\nonumber\\ 
&-& \frac{m_s^2}{2\pi^2} \int{\left[f_{\nu_s}(p)+
f_{\bar\nu_s}(p)\right] \left[(p^2+m_s^2)/T^2\right]^{-1/2} p^2
dp}.\nonumber
\end{eqnarray}
This, together with the energy density and pressure in the sterile
neutrinos and antineutrinos calculated from their time-dependent
distribution functions, allows one to readily arrive at a consistent
time-temperature evolution.

\end{document}